\documentclass[a4paper,11pt,final]{article}
\pdfoutput=1 

\usepackage{jheppub} 

\usepackage[T1]{fontenc} 
\usepackage[utf8]{inputenc}
\usepackage{easy-todo}
\usepackage{mathtools} 
\usepackage{xifthen}
\usepackage{siunitx} 
\usepackage{tensor}
\usepackage{slashed}
\usepackage{cancel}
\usepackage{nicefrac}
\usepackage{subcaption}
\usepackage[noabbrev]{cleveref}
\crefname{equation}{eq.}{eqs.}%
\Crefname{equation}{Equation}{Equations}%
\usepackage[normalem]{ulem}
\usepackage{enumitem}
\setlist{noitemsep}

\title{{\Large Dirac vs.\@ Majorana HNLs (and their oscillations) at SHiP}}

\author[a]{J.-L. Tastet}
\author[b]{I. Timiryasov}

\affiliation[a]{Discovery Center, Niels Bohr Institute, University of Copenhagen,\\
  Blegdamsvej 17, DK-2010, Copenhagen, Denmark}
\affiliation[b]{Institute of Physics, Laboratory for Particle Physics and Cosmology,\\
  École Polytechnique Fédérale de Lausanne, CH-1015 Lausanne, Switzerland}

\emailAdd{jeanloup@nbi.ku.dk}
\emailAdd{Inar.Timiryasov@epfl.ch}


\arxivnumber{1912.05520}

\abstract{%
  SHiP is a proposed high-intensity beam dump experiment set to operate at the CERN SPS.
  It is expected to have an unprecedented sensitivity to a variety of models containing feebly interacting particles,
  such as Heavy Neutral Leptons (HNLs).
  Two HNLs or more could successfully explain the observed neutrino masses through the seesaw mechanism.
  If, in addition, they are quasi-degenerate, they could be responsible for the baryon asymmetry of the Universe.
  Depending on their mass splitting, HNLs can have very different phenomenologies:
  they can behave as Majorana fermions --- with lepton number violating~(LNV) signatures,
  such as same-sign dilepton decays --- or as Dirac fermions with only lepton number conserving~(LNC) signatures.
  In this work, we quantitatively demonstrate that LNV processes can be distinguished from LNC ones at SHiP,
  using only the angular distribution of the HNL decay products.
  Accounting for spin correlations in the simulation and using boosted decision trees for discrimination,
  we show that SHiP will be able to distinguish Majorana-like and Dirac-like HNLs
  in a significant fraction of the currently unconstrained parameter space.
  If the mass splitting is of order $10^{-6}\;\si{eV}$, SHiP could even be capable of resolving HNL oscillations,
  thus providing a direct measurement of the mass splitting.
  This analysis highlights the potential of SHiP to not only search for feebly interacting particles,
  but also perform model selection.
}

\begin{document}

\maketitle
\flushbottom

\renewcommand{\vec}[1]{\ensuremath{\mathbf{#1}}}
\providecommand{\gvec}[1]{\ensuremath{\pmb{#1}}}
\providecommand{\vhat}[1]{\ensuremath{\hat{\mathbf{#1}}}}
\providecommand{\op}[1]{\ensuremath{\hat{#1}}}
\providecommand{\id}{\ensuremath{\mathbb{1}}}
\renewcommand{\Re}{\ensuremath{\operatorname{Re}}}
\renewcommand{\Im}{\ensuremath{\operatorname{Im}}}
\renewcommand{\det}[1]{\ensuremath{\mathrm{det}\!\left(#1\right)}}
\providecommand{\tr}[1]{\ensuremath{\mathrm{tr}\!\left(#1\right)}}
\providecommand{\diag}[1]{\ensuremath{\mathrm{diag}\left(#1\right)}}
\providecommand{\vket}[1]{\ensuremath{\left| #1 \right\rangle}}
\providecommand{\vbra}[1]{\ensuremath{\left\langle #1 \right|}}
\providecommand{\vbraket}[1]{\ensuremath{\left\langle #1 \right\rangle}}
\providecommand{\fket}[2]{\ensuremath{\expandafter\csname#1\endcsname| #2 \expandafter\csname#1\endcsname\rangle}}
\providecommand{\fbra}[2]{\ensuremath{\expandafter\csname#1\endcsname\langle #2 \expandafter\csname#1\endcsname|}}
\providecommand{\fbraket}[2]{\ensuremath{\expandafter\csname#1\endcsname\langle #2 \expandafter\csname#1\endcsname\rangle}}
\providecommand{\ket}[2][]{\ifthenelse{\isempty{#1}}{\vket{#2}}{\fket{#1}{#2}}}
\providecommand{\bra}[2][]{\ifthenelse{\isempty{#1}}{\vbra{#2}}{\fbra{#1}{#2}}}
\providecommand{\braket}[2][]{\ifthenelse{\isempty{#1}}{\vbraket{#2}}{\fbraket{#1}{#2}}}
\providecommand{\abs}[1]{\ensuremath{\left| #1 \right|}}
\providecommand{\norm}[1]{\ensuremath{\left\| #1 \right\|}}
\providecommand{\del}{\ensuremath{\nabla}}
\providecommand{\dd}{\ensuremath{\partial}}
\providecommand{\dslash}{\ensuremath{\slashed{\partial}}}
\providecommand{\Dslash}{\ensuremath{\slashed{D}}}
\providecommand{\cc}{\ensuremath{\mathrm{c.c.}}}
\providecommand{\hc}{\ensuremath{\mathrm{h.c.}}}
\providecommand{\dc}[1]{\ensuremath{\overline{#1}}} 
\renewcommand{\d}[2][]{\ifthenelse{\isempty{#1}}{\ensuremath{\!\mathrm{d}#2\:}}{\ensuremath{\!\mathrm{d}^{\mathrm{#1}}\!#2\:}}}
\providecommand{\D}[1]{\ensuremath{\left[D #1\right]\:}}
\providecommand{\diff}{\ensuremath{\mathrm{d}}}
\sisetup{exponent-product = \cdot, output-product = \cdot}
\providecommand{\unit}[1]{\,\si{#1}}
\providecommand\eqdef{\ensuremath{\mathrel{\overset{\makebox[0pt]{\mbox{\normalfont\tiny\sffamily def}}}{=}}}}
\providecommand{\ie}{i.e.\@}
\providecommand{\eg}{e.g.\@}
\providecommand{\cf}{\textit{cf.\@}}
\providecommand{\Cf}{\textit{Cf.\@}}
\providecommand{\fig}{fig.\@}

\hyphenation{pseudo-scalar}

\providecommand{\nuMSM}{\ensuremath{\mathrm{\nu MSM}}}
\providecommand{\ship}{SHiP}
\providecommand{\dLips}{\ensuremath{\mathrm{dLips}\,}}
\providecommand{\U}{\ensuremath{\mathcal{U}}}
\providecommand{\vev}{\ensuremath{\braket{\abs{\phi}}}}


\section{Introduction}
\label{sec:introduction}

The experimentally observed non-vanishing neutrino mass differences are among
a few firmly established deviations from the Standard Model~(SM) predictions.
An economic way of generating the light neutrino masses is to introduce
heavy singlet fermions with Majorana mass terms into the model%
~\cite{Minkowski:1977sc,GellMann:1980vs,Mohapatra:1979ia,Yanagida:1980xy,Schechter:1980gr,Schechter:1981cv}.
The masses of the active neutrinos in this extension of the SM are determined by the type-I seesaw formula and
at least two singlet fermions are needed to accommodate the two observed mass differences of light neutrinos.
A consequence of this mechanism is the presence of heavy Majorana fermions which mix with active neutrinos.
The mass scale of these Majorana fermions --- Heavy Neutral Leptons (HNLs) --- is not fixed.
It can be below the electroweak scale,%
\footnote{%
  An argument in favour of the low-scale seesaw comes from the measured values of the Higgs and top masses.
  HNLs with masses below the electroweak scale are not destabilising the Higgs mass~\cite{Vissani:1997ys,Bezrukov:2012sa}.
}
like in the \nuMSM{}~\cite{Asaka:2005an,Asaka:2005pn},
where two HNLs are responsible for the light neutrino masses
and generating the Baryon Asymmetry of the Universe (BAU)
via $CP$-violating oscillations during their production.

From the FIP (feebly interacting particles) search point of view,
HNLs with masses below that of a $B$~meson are the most accessible in the foreseeable future~\cite{pbc_2019}.
There is a vast program to search for HNLs at \emph{intensity frontier} experiments,
either LHC-based, such as MATHUSLA~\cite{chou_new_2017,curtin_long-lived_2018,Alpigiani:2018fgd},
FASER~\cite{faser_2018,Kling:2018wct,Ariga:2018zuc}, CODEX-b~\cite{gligorov_searching_2018,Aielli:2019ivi},
AL3X~\cite{gligorov_leveraging_2018,Dercks:2018wum} and ANUBIS~\cite{Bauer:2019vqk},
or at beam-dump facilities, such as DUNE~\cite{Akiri:2011dv,Krasnov:2019kdc,ballett_heavy_2019} (using the near detector),
NA62\textsuperscript{++}~\cite{NA62:2017rwk,drewes_na62_2018} (in dump mode)
and \ship{}~\cite{ship_collaboration_facility_2015,alekhin_facility_2016,ship_collaboration_sensitivity_2018}.
Comparative studies of the \emph{exclusion} limits expected from these experiments
have been performed in refs.~\cite{Helo:2018qej,Boiarska:2019jcw,Bondarenko:2019yob,Chun:2019nwi}.
If a candidate HNL signal were to be observed, the latter three experiments would be sensitive
to both its mass and mixing angles.

\begin{figure}
  \centering
  \includegraphics[width=\textwidth]{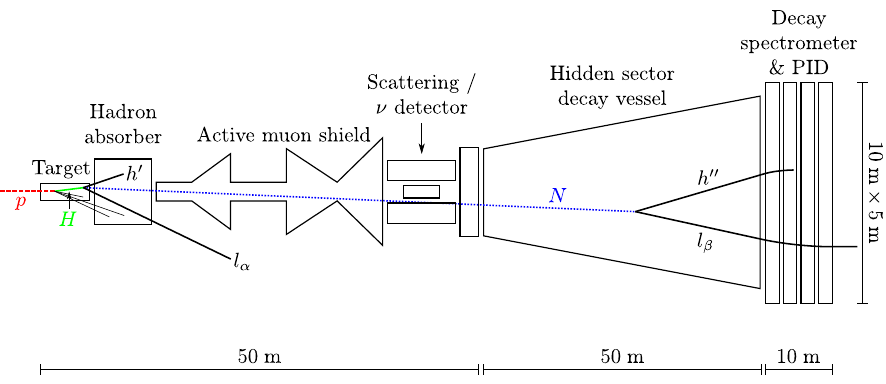}
  \caption{Sketch of the \ship{} experiment, with the decay chain $H \to h' l_{\alpha} (N \to l_{\beta} h'')$.}
  \label{fig:ship}
\end{figure}

\ship{} is a proposed beam-dump experiment (represented in \cref{fig:ship}) set to operate at the CERN SPS.
It will use an intense, \SI{400}{GeV} proton beam from the SPS,
dumped on a thick target in order to produce a large number of heavy hadrons,
which subsequently decay into Standard Model (SM) or feebly-interacting particles.
\ship{} is designed to provide a background-free environment to look for the decays of these heavy FIPs.
To this end, a hadron absorber located right after the target absorbs most SM particles.
It is followed by an active muon shield which deflects the muons away from the experimental cavern.
The main detector consists of a decay volume %
--- evacuated in order to reduce the neutrino background, and surrounded by vetos --- %
with a tracker and a calorimeter located at its far end, enabling it to reconstruct the decay event.

In order to generate the light neutrino masses via the seesaw mechanism, HNLs must be Majorana fermions,
which violate the total lepton number.
However, if the mass splitting is small enough, they can pair to form a coherent superposition of two quasi-degenerate Majorana fermions,
which behaves almost like a Dirac fermion. Such a combination is dubbed ``\emph{quasi-Dirac pair}''.
In this case, the mixing angles can exceed the naive seesaw limit $U^2 \approx m_{\nu}/M_N$
\cite{Shaposhnikov:2006nn,Kersten:2007vk,Moffat:2017feq},
where $m_{\nu}$ and $M_N$ are respectively the mass scales of light neutrinos and HNLs.
This is possible because a quasi-Dirac fermion approximately conserves the total lepton number,
hence protecting the light neutrino masses.
For instance, the \nuMSM{}~\cite{Asaka:2005an,Asaka:2005pn} contains such a quasi-Dirac pair if one requires the mass degeneracy which is needed for baryogenesis~\cite{Akhmedov:1998qx,Asaka:2005pn} and especially for late-time leptogenesis~\cite{Canetti:2012kh}. Quasi-Dirac pairs also naturally appear
in some models of neutrino mass generation,
such as the inverse seesaw~\cite{Mohapatra:1986aw,Mohapatra:1986bd}
and the linear seesaw~\cite{Akhmedov:1995ip,Akhmedov:1995vm}.
This near degeneracy of the HNL masses leads to coherent HNL oscillations.
In the \nuMSM{}, these oscillations in the early Universe are responsible for baryogenesis.

For sufficiently light ($\lesssim \SI{10}{GeV}$) HNLs like the ones accessible at \ship{},
LNV may be experimentally observable even when they form a quasi-Dirac pair~\cite{anamiati_quasi-dirac_2016,drewes_lepton_2019}.
We can distinguish three cases,%
\footnote{%
  To be generic, we have included the more exotic cases of a single Dirac or Majorana HNL.
  The limits presented below are for a quasi-Dirac pair,
  which only differs from those in the number of events produced.
}
depending on the scale of the oscillation phase~$\delta M \tau$,
where $\delta M$ is the mass splitting of the quasi-Dirac pair and $\tau$ the typical proper time probed:
\begin{enumerate}
\item \textbf{Dirac-like HNL:} One Dirac HNL or a quasi-Dirac pair with an oscillation period exceeding
  the HNL lifetime or detector size ($\delta M \tau \ll 2\pi$).%
  \footnote{As pointed out in ref.~\cite{drewes_lepton_2019}, for most experiments,
    this possibility might be technically unnatural due to the very small mass splitting
    needed to satisfy the inequality.}
  Only LNC processes can be observed. 
\item \textbf{Majorana-like HNL:} One Majorana HNL or a quasi-Dirac pair with a lifetime and detector size exceeding
  the oscillation period ($\delta M \tau \gg 2\pi$).
  Both LNC and LNV processes can be observed, with equal integrated rates (see \cref{sec:oscillations}).
\item \textbf{Manifestly quasi-Dirac HNLs:}
  An interesting case occurs when the oscillation period is comparable to
  the HNL lifetime or to the size of the detector%
  \footnote{%
    Interestingly, the mass difference needed to generate DM in the \nuMSM{},
    as found in ref.~\cite{Canetti:2012kh}, is exactly in this borderline range.
  }
  ($\delta M \tau \sim 2\pi$):
  the experiment may then be sensitive to the coherent oscillations of HNLs.
\end{enumerate}

If HNLs were to be observed at \ship{}, the detection or non-observation of lepton number violation
and HNL oscillations would allow constraining models and their parameters.
The most relevant LNV process at \ship{} is the well-studied \emph{same-sign dilepton decay}:
$H\to [h'] l_{\alpha}^+ (N \to h'' l_{\beta}^+)$, where $H$, $h'$ and $h''$ are hadrons (with $h'$ possibly missing),
and $l_{\alpha}^+$, $l_{\beta}^+$, $\alpha, \beta = e, \mu, \tau$ are charged leptons of potentially different generations.
Due to suppressed background, this type of signature is a smoking gun for HNLs in accelerator searches. 
However, at beam-dump experiments,
the heavy hadron decay which produces the HNL takes place inside the target,
and therefore the charge of the primary lepton $l_{\alpha}$ cannot be observed.
Naively, it seems that the information about the HNL production is lost,
since the charge of the secondary lepton $l_{\beta}$, by itself, is not enough to tell apart LNC and LNV processes.
As we shall see in this paper, it turns out that the HNL decay products nevertheless carry important information.
Namely, their distribution is different for LNC and LNV processes. 
Not only does this allow distinguishing Majorana-like from Dirac-like HNLs given sufficiently many events,
but the knowledge of these distributions
can also be used to \emph{resolve} HNL oscillations and directly measure the mass splitting.

Estimating these two distributions is complicated by the presence of
a variety of two- and three-body production channels.
In addition, the parent hadrons are produced with a finite spectrum.
As we shall see in \cref{sec:lab_frame_distribution}, this smears the distributions, making them look more similar.
Therefore, in order to assess whether SHiP will be able to discriminate between Majorana- and Dirac-like HNLs,
an accurate treatment of all production channels, including spin correlations, is required.
This is accomplished using a Monte-Carlo simulation.

The angular distribution of HNL decay products has been studied in a collider setting
for decays which are not fully reconstructible~\cite{arbelaez_probing_2018,dib_signatures_2017,1810.07210}
(such as trilepton decays),
as well as for beam-dump experiments~\cite{cvetic_probing_2012,balantekin_addressing_2019}.
Our analysis improves on the latter by not relying on HNLs being produced as helicity eigenstates,
by handling a larger class of production channels,
by considering the full phase-space distribution of the HNL decay products (instead of just their energy)
and by producing a concrete sensitivity estimate using a realistic geometry and heavy meson spectrum for \ship{}.

This paper is organized as follows.
In \cref{sec:model}, we review the Standard Model extended with HNLs,
and discuss lepton number violation and coherent HNL oscillations.
In \cref{sec:lnv_at_ship}, we analyze the different signatures of LNC and LNV processes at the \ship{} experiment.
In \cref{sec:simulation_and_analysis}, we propose a strategy to detect LNV and reconstruct HNL oscillations.
Finally, in \cref{sec:results}, we present the sensitivity of \ship{} to LNV achieved through this method,
as well as a possible signature of HNL oscillations.
Technical details about the simulation and the statistical analysis are respectively provided in \cref{sec:simulation,sec:classification}.

\section{Model}
\label{sec:model}

\subsection{Heavy Neutral Leptons}
\label{sec:hnls}

We consider the Standard Model extended with $\mathcal{N}$ HNLs $N_{I}$, which are spin-$\frac{1}{2}$ SM singlets with Majorana masses $M_I$,
and new Yukawa couplings $Y_{\alpha I}^{\nu}$, with $\alpha=e,\mu,\tau$ the lepton flavor index.
Using the conventions from ref.~\cite{dreiner_two-component_2010}:
\begin{equation}
    \mathcal{L} = \mathcal{L}_{\mathrm{SM}} + \frac{i}{2} N_I^{\dagger} (\bar{\sigma}\cdot \dd) N_I
    - (Y_{\alpha I}^{\nu})^* (\phi \cdot L_{\alpha}) N_I - \frac{M_I}{2} N_I N_I + \hc
\end{equation}
After electroweak symmetry breaking,
the Yukawa interaction generates a Dirac mass term $(m_D)_{\alpha I} = \frac{v}{\sqrt{2}} (Y_{\alpha I}^{\nu})^*$,
resulting in a non-diagonal, symmetric Dirac-Majorana mass term for neutrinos~\cite{Giunti:2007ry}:
\begin{equation}
  \!M_{\mathrm{DM}} = -\frac{1}{2} \begin{pmatrix} \nu^T & N^T \end{pmatrix} \begin{pmatrix} 0 & m_D^T \\ m_D & M_M \end{pmatrix} \begin{pmatrix} \nu \\ N \end{pmatrix} + \hc
\end{equation}
where $M_M = \diag{M_I\dots}$.
Using a unitary transformation of the fields (Takagi factorization~\cite{takagi_1927}),
the mass matrix can be brought to a diagonal form:
\begin{align}
  &\nu_{\alpha} = U_{\alpha i} n_i \quad \text{ and } \quad N_I = U_{I i} n_i \\
  &M_{\mathrm{DM}} = -\frac{m_i}{2} (n_i n_i + n_i^{\dagger} n_i^{\dagger})
\end{align}
In the limit $|M_M| \gg |m_D|$, we can use an approximate block factorization,
leading to the mass eigenstates $n_i \cong \nu_i, N_I$ mixing with the flavor fields as:
\begin{align}
  \nu_{\alpha} &\cong U_{\alpha i}^{\mathrm{PMNS}} \nu_i + \Theta_{\alpha I} N_I \label{eq:mixing} \\
  \Theta_{\alpha I} &\cong M_I^{-1} (m_D)_{\alpha I} \label{eq:mixing_angles}
\end{align}
and the following mass sub-matrices:
\begin{align}
  m_{\alpha \beta} &\cong - \sum_I \frac{(m_D)_{\alpha I} (m_D)_{\beta I}}{M_I}
  \cong - \sum_I M_I \Theta_{\alpha I} \Theta_{\beta I} \label{eq:seesaw} \\
  m_{IJ} &\cong M_I \delta_{IJ}
\end{align}
The choice of the mass scale~$M_M$ and Yukawa couplings~$Y_{\alpha I}^{\nu}$
is not uniquely dictated by low-energy neutrino observables, and should be fixed otherwise.

\begin{figure}
  \centering
  \begin{subfigure}{0.49\linewidth}
    \centering
    \includegraphics[width=\linewidth]{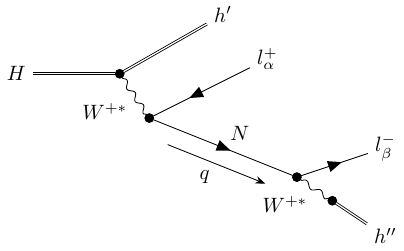}
    \caption{LNC}
    \label{fig:lnc_chain}
  \end{subfigure}
  \begin{subfigure}{0.49\linewidth}
    \centering
    \includegraphics[width=\linewidth]{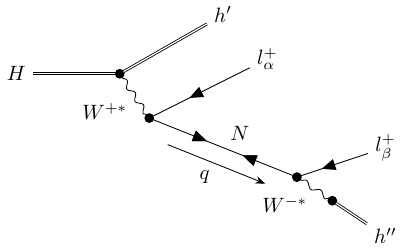}
    \caption{LNV}
    \label{fig:lnv_chain}
  \end{subfigure}
  \caption{Lepton number conserving and violating decay chains for $H \rightarrow h' l_{\alpha} (N \rightarrow l_{\beta} h'')$.}
  \label{fig:lnc_lnv_chains}
\end{figure}

The Standard Model features an accidental symmetry --- lepton number --- %
which, at tree level, is conserved for massless or Dirac neutrinos,
but is violated by the Majorana mass term of HNLs.
Charged leptons and neutrinos have lepton number~$+1$,
while charged anti-leptons and anti-neutrinos have lepton number~$-1$.
If lepton number is conserved (LNC),
then the only allowed Feynman diagrams are those with a conserved flow of lepton number
(represented by the arrow on the fermion lines of leptons),
like the opposite-sign dilepton decay of a heavy hadron shown in \cref{fig:lnc_chain}.
On the other hand, in the presence of lepton number violating (LNV) operators,
processes like the same-sign dilepton decay shown in \cref{fig:lnv_chain} become possible.
Lepton number violation can also manifest itself in neutral-current processes or in neutrinoless double-$\beta$ decay.
Whether such LNV transitions actually happen depends on the specific model.

In the past decade, a class of low-scale seesaw models have risen in popularity,
such as the \nuMSM{}~\cite{Asaka:2005pn},
not least because of their falsifiability at existing or proposed experiments.
In these models, $M_M$ is postulated to be below the electroweak scale.
The seesaw formula~\eqref{eq:seesaw} requires at least $2$~HNLs to explain the two observed mass differences.
If their parameters are arbitrary, then the smallness of the light neutrino masses is achieved
through small Yukawa couplings of order $Y^{\nu} \sim \frac{1}{v}\sqrt{|m_{\nu}| |M_M|}$,
leading to squared mixing angles $|\Theta|^2 \sim |m_{\nu}| / |M_M|$.
For a typical HNL with $M_M \sim \SI{1}{GeV}$, this gives $|\Theta|^2 \sim 10^{-11}$,
a number that is too small to be probed at any current or proposed experiment.

However, multiple HNLs can have mixing angles well above the seesaw limit,
yet at the same time produce the correct neutrino masses in a technically natural way,
if a certain symmetry is imposed on their Yukawa couplings.
If we consider for simplicity $\mathcal{N}=2$ nearly degenerate HNLs $N_{1,2}$, their mixing angles should be related by
$\Theta_{\alpha 2} \approx \pm i \Theta_{\alpha 1}$~\cite{Shaposhnikov:2006nn,Kersten:2007vk}.
Such HNLs form a quasi-Dirac fermion, which approximately conserves the total lepton number.
This implies that the usual searches for naive LNV effects (\eg{} same-sign dilepton decays),
may return null results even if HNLs \emph{are there}.

Below we discuss an important consequence of the approximate nature of this lepton number conservation:
HNL oscillations, and how quasi-Dirac HNLs can phenomenologically behave either as Majorana or Dirac HNLs
depending on their mass splitting~$\delta M$ and the length scale probed at the experiment.

\subsection{Coherent oscillations of Heavy Neutral Leptons}
\label{sec:oscillations}

The \ship{} experiment is only sensitive to \si{GeV}-scale HNLs,
with mixing angles significantly above the seesaw limit~\cite{ship_collaboration_sensitivity_2018}.
Therefore it can only probe the quasi-Dirac regime described above.
Apart from a small mass splitting~$\delta M \ll M$, the two HNLs are otherwise identical.
Since these two HNLs cannot be distinguished in any realistic experiment,
they both mediate the same processes and each contribute to the total transition amplitude,
resulting in interference.
Only the initial and final-state particles, which strongly interact with the environment,
are measured in the quantum mechanical sense.
In order to accurately describe processes involving multiple HNLs,
it is therefore necessary to consider them as intermediate particles
within a larger process consisting of the HNL production, propagation and decay,
and only square the overall transition amplitude between the observed, external particles.
This can be formulated rigorously within the framework of the external wave packet model%
~\cite{sachs_interference_1963, giunti_treatment_1993}
(see also~\cite{beuthe_oscillations_2003,akhmedov_neutrino_2010,akhmedov_neutrino_2011,akhmedov_quantum_2019}
and references therein for recent reviews).
Let us note in passing that this description automatically takes care of spin correlations between
the particles taking part in the HNL production and decay.

In what follows, we consider a typical reconstructible decay chain at \ship{},
as depicted in \cref{fig:lnc_lnv_chains}.
We will postpone the detailed discussion of this process to \cref{sec:lnv_at_ship}.
A heavy hadron~$H$ produced in the target decays at space-time coordinates~$x_P$
into an HNL~$N_I$, a charged lepton~$l_{\alpha}$ (the \emph{primary} lepton),
and an optional hadron~$h'$.
If the HNL is sufficiently long-lived, it can propagate a macroscopic distance before
decaying at~$x_D$ into a charged lepton~$l_{\beta}$ (the \emph{secondary} lepton) and a hadron~$h''$.

The slightly different masses of the HNLs mediating the process lead to different dispersion relations $q_I^2 = M_I^2$.
As a consequence,
the space-time-dependent phase $e^{-iq_I\cdot (x_D-x_P)}$ acquired by the HNL between its production and decay
will differ slightly for each mass eigenstate.
When squaring the amplitude in order to obtain the differential decay rate,
the interference terms between the partial amplitudes coming from different mass eigenstates
will therefore feature a space-time-dependent modulation: HNL oscillations.
The external wave packet model allows one to unambiguously establish the expression for the oscillation phase
and check that the entire process remains coherent in all experimentally relevant situations.

The present paper does not aim to be a detailed study of HNL oscillations,
which have already been covered in various settings and limits in the literature%
~\cite{Asaka:2005pn,eijima_parameter_2018,1409.4265,1505.04749,anamiati_quasi-dirac_2016,1709.03797,Das:2017hmg,1805.00070,1810.07210}.
Therefore, we will only quote the main result.
Let $\diff\hat{\Gamma}_{\alpha\beta}^{\pm\pm}$ be the differential rate for
the above-described process $H \rightarrow [h'] l_{\alpha}^{\pm} (N \rightarrow l_{\beta}^{\pm} h'')$
mediated by a single Majorana HNL~$N$,
in the (unphysical) limit of a unit mixing angle between the HNL and
the active flavor $\alpha$ at its production vertex, with flavor $\beta$ at its decay vertex,
and without the absorptive part.
The coherent differential rate $\smash{\diff\Gamma_{\alpha\beta}^{\pm\pm}(\tau)}$
in the presence of $\mathcal{N}$~nearly degenerate HNLs mediating the process,
as a function of the proper time $\tau = \sqrt{(x_D-x_P)^2}$ between the HNL production and decay vertex, is then:
\begin{equation}
  \diff\Gamma_{\alpha\beta}^{\pm\pm}(\tau) =
  \abs{\sum_{I=1}^{\mathcal{N}} \Theta_{\alpha I}^{\pm} \Theta_{\beta I}^{\pm} e^{-i M_I \tau - \frac{\Gamma_I}{2} \tau}}^2
  \diff\hat{\Gamma}_{\alpha\beta}^{\pm\pm}
  \label{eq:coherent_width}
\end{equation}
where $M_I$ is the (Majorana) mass of the $I$-th heavy mass eigenstate, $\Gamma_I$ its total width,
and we have used the shorthand notation $\Theta^+ \eqdef \Theta^*$ and $\Theta^- \eqdef \Theta$.

In the case of $\mathcal{N}=2$ HNLs forming a quasi-Dirac pair,
\ie{} $M_1 = M - \frac{\delta M}{2}$, $M_2 = M + \frac{\delta M}{2}$,
$\Theta_{\alpha 2} \cong \pm i\Theta_{\alpha 1}$ and $\Gamma_1 \cong \Gamma_2 \eqdef \Gamma$,
the coherent differential rate reduces to:
\begin{equation}
  \diff\Gamma_{\alpha\beta}^{\pm\pm}(\tau) \cong
  2 \abs{\Theta_{\alpha 1}}^2 \abs{\Theta_{\beta 1}}^2 \left( 1 \pm \cos\left(\delta M \tau\right) \right) e^{-\Gamma \tau}
  \diff\hat{\Gamma}_{\alpha\beta}^{\pm\pm}
  \label{eq:quasi_dirac}
\end{equation}
where the $+$ sign is for lepton number conserving processes
($\diff\Gamma_{\alpha\beta}^{+-}$ and $\diff\Gamma_{\alpha\beta}^{-+}$),
and the $-$ sign for lepton number violating ones
($\diff\Gamma_{\alpha\beta}^{++}$ and $\diff\Gamma_{\alpha\beta}^{--}$).
Notice how in the quasi-Dirac limit, the oscillation pattern does not explicitly depend
on the lepton flavors $\alpha$ and $\beta$, but only on whether the process is LNC or LNV.
If $\delta M$ vanishes exactly, HNLs form a Dirac fermion and LNV effects are completely absent.
Recently, $CP$-violating HNL oscillations have attracted some interest%
~\cite{1904.04787,1905.03097,1906.09470,abada_interference_2019}.
However, here we can see that $CP$-violation is suppressed in the quasi-Dirac limit.

Throughout this paper, we will focus on the case where $\Gamma \tau \ll 1$,
which is the most relevant for \ship{},
and drop the exponentially decaying factor.
Analysing formula~\eqref{eq:quasi_dirac}, we see that there are three regimes of interest,
depending on the mass splitting $\delta M$ and proper time scale~$\tau$ probed at the experiment:
\begin{itemize}
\item If $\delta M \tau \ll 2\pi$, the HNL pair is observed before the onset of oscillations,
  and it behaves like a single Dirac HNL, \ie{} we cannot observe lepton number violation.
\item If $\delta M \tau \gg 2\pi$, fast oscillations are averaged out,
  and the HNL pair behaves like a single Majorana HNL,
  with equal integrated decay rates for LNC and LNV channels.%
  \footnote{
    In the rest frame of a single on-shell, Majorana HNL,
    the only ``memory'' of the production process is the HNL spin.
    To perform the phase-space integration for the HNL decay,
    one can always choose a frame where the HNL is at rest and with a fixed spin projection,
    hence resulting in the same integrated rates for LNC and LNV processes.
  }
\item If $\delta M \tau \sim 2\pi$, oscillations must be accounted for.
  If it is possible to experimentally reconstruct, for each selected event,
  the proper time~$\tau$ between the production and decay vertex of the HNL,
  then oscillations can be resolved,
  \ie{} the $\tau$-differential event rates for LNC / LNV will show a periodic modulation according to \cref{eq:quasi_dirac}.
\end{itemize}
At \ship{}, the proper time scale $\tau$ is about $\SI{2}{m}$ for sufficiently long-lived HNLs.
It corresponds to the average time between the production and decay of an \emph{observed} HNL, in its rest frame.
Therefore, the critical mass splitting separating the three regimes %
--- near which oscillations are resolvable --- is about $10^{-6}\;\si{eV}$.

\section{Probing lepton number violation at \ship{}}
\label{sec:lnv_at_ship}

Many collider searches for Majorana HNLs%
~\cite{lhcb_collaboration_search_2014,atlas_collaboration_search_2015,cms_collaboration_search_2015,cms_collaboration_search_2016}
are sensitive to lepton number violation through the charges of the leptons
produced at the HNL production and decay vertex.
Indeed, due to the chiral nature of the weak interaction,
they unambiguously tell the chiral projection through which the HNL interacts at a given vertex.
In theory, a same-sign dilepton decay (either prompt or displaced)
would thus provide clear evidence for lepton number violation
(although, in practice, significant standard model backgrounds exist for prompt decays).

At \ship{}, similar numbers of mesons and anti-mesons are expected to be produced.%
\footnote{%
  Unless cascade production significantly alters the results from ref.~\cite{lebc_d-meson_1987}.
  The charm spectrum will be measured at \ship{} prior to data taking~\cite{ship_collaboration_measurement_2017}.
  Asymmetries, if present, can only improve the classification accuracy,
  since the secondary lepton charge would then carry \emph{some} information.
}
This leads to similar numbers of HNLs being produced along with positively and negatively charged primary leptons.
Consequently, the secondary lepton charge contains very little information
as to whether the process is LNC or LNV.
To lift this degeneracy, it becomes necessary to look at new observables.

\begin{figure}[htb!]
  \centering
  \includegraphics[width=\textwidth]{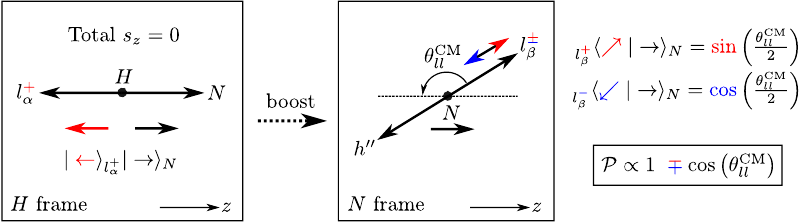}
  \caption{%
    This sketch explains the origin of the different angular correlations for LNC and LNV processes.
    For simplicity, here we consider two-body primary and secondary decays involving only pseudoscalar mesons,
    and the masses of the charged leptons and of $h''$ are neglected.
    For definiteness, the charge of the primary lepton %
    --- which is produced inside the target and thus inaccessible --- is also fixed to~$+$.
    Since the HNL is a Majorana fermion, the secondary lepton $l_{\beta}$ can have either charge.
    However, due to angular momentum conservation, the lepton~$l^+_\alpha$ and the HNL~$N$ are produced
    with opposite spin projections in the rest frame of the heavy meson $H$.
    Because of chiral suppression (which is more effective for light fermions),
    the charge of the primary lepton is correlated with its spin
    (\eg{} in the massless limit, $l_{\alpha}^+$ has helicity $+\frac{1}{2}$) and hence with the HNL spin.
    For the same reason, the angular distribution of the decay products of the resulting HNL spin eigenstate
    (which is unaffected by a boost along the quantization axis)
    will therefore depend on the secondary lepton charge.
    The very same formula for the probability $\mathcal{P}$ also holds for $CP$-conjugated channels,
    with the $+$ sign for LNC and the $-$ sign for LNV.
    The general case (massive, with two- or three-body primary decay) is discussed in \cref{sec:kinematics}.
  }
  \label{fig:sketch}
\end{figure}

Luckily, the HNL lepton number is not the only quantum number conserved by the weak interaction.
The HNL also carries spin~$\frac{1}{2}$,
and the total angular momentum is always conserved.
When the HNL is produced, its spin is correlated (opposite if $H$ and $h'$ are pseudoscalar) with that of the primary lepton.
Due to chiral suppression,
the spin of the primary lepton is itself correlated with its lepton number
(see for example the left part of \cref{fig:sketch}).
This suggests that by looking at the angular distribution of the secondary particles --- which may be observable --- %
we should be able to obtain information about the primary interaction,
and thus whether the process was LNC or LNV (see the right part of \cref{fig:sketch}).
This realization was the starting point of the present work.
More generally, we expect LNC and LNV decay chains to have different kinematics due to their different Lorentz structures,
potentially allowing us to distinguish them without directly observing the primary decay.

In \cref{sec:hnl_production_decay}, we describe the relevant HNL production and decay channels at \ship{};
in \cref{sec:kinematics}, we quantitatively compare the angular distributions for LNC and LNV processes,
and in \cref{sec:lab_frame_distribution} we discuss how this affects the observable momenta in a beam-dump setting.

\subsection{HNL production and decay at \ship{}}
\label{sec:hnl_production_decay}

At \ship{}, most HNLs are produced in heavy meson decays through flavor-changing charged currents,
as discussed in ref.~\cite{bondarenko_phenomenology_2018}.
In addition, for the present analysis, we will only consider fully reconstructible HNL decays
such as $N \rightarrow l_{\beta}^{\mp} \pi^{\pm}$,
producing only charged particles which are sufficiently long-lived to be detected
by the tracking station located at the end of the decay vessel.
Those are also mediated by the charged-current interaction.

Without losing generality, we can therefore consider the generic lepton number conserving and violating processes
$H \rightarrow [h'] l_{\alpha} (N \rightarrow l_{\beta} h'')$
represented in \cref{fig:lnc_chain,fig:lnv_chain}, respectively,
as well as their $CP$-conjugates.
$H$ denotes a heavy hadron (typically a $D_{[s]}$ or $B_{[c]}$~meson at \ship{}),
$h'$ and $h''$ are hadrons (with $h'$ missing for two-body primary decays),
and $l_{\alpha}^{\pm}$ and $l_{\beta}^{\pm}$ are respectively the primary and secondary leptons.

Since the heavy hadron~$H$ is typically short-lived, the primary decay takes place inside the target and cannot be observed.
If the HNL is sufficiently long-lived (we will assume this to be the case throughout this paper),
it can propagate a macroscopic distance before decaying,
and leave a very displaced vertex inside the \ship{} decay vessel.
For the selected decay channels $N \rightarrow l_{\beta}^{\mp} \pi^{\pm}$, this secondary vertex can be fully reconstructed.

In the present study, we will restrict ourselves to HNL masses between the $K$ and $D_s$ thresholds.
Masses below the $K$ threshold have already been heavily constrained~\cite{pbc_2019},
while above the $D_s$ mass, HNLs are mainly produced in $B$~meson decays,
whose spectrum cannot be directly measured at the beam dump,
making our analysis more sensitive to modeling errors.

\subsection{Angular correlations in LNC and LNV decay chains}
\label{sec:kinematics}

In order to study the angular correlations between all final-state particles,
spin correlations between the primary and secondary decay must be accounted for.
Those result from the non-observation of the HNL spin,
which leads to interference between the two spin eigenstates $N_s$, $s=\pm \frac{1}{2}$
(similarly to how the non-observation of its precise mass allows for flavor oscillations).
To compute the overall transition amplitude, we can therefore use the same trick as for oscillations,
\ie{} treat the primary and secondary decays as a single process.

To simplify the calculations, in this section we will focus on the case of a single Majorana HNL,
which mediates both LNC and LNV decay chains with equal rates,
and we will omit the absorptive part of the amplitude
(\ie{} we will study $\diff\hat{\Gamma}_{\alpha\beta}^{\pm\pm}$ instead of $\diff\Gamma_{\alpha\beta}^{\pm\pm}(\tau)$).
We do not lose generality in doing so,
because the effect of multiple nearly degenerate HNLs and their finite lifetime can be factored out,
and subsequently recovered, using \cref{eq:coherent_width,eq:quasi_dirac}.
To keep the notation light, we will from now on drop the HNL index~$I=1$.

Since we are only concerned with long-lived HNLs,
which are produced on their mass shell and have well separated, localized production and decay vertices,
the momentum~$q$ of the HNL is practically fixed, which allows factorizing the transition amplitude as:
\begin{equation}
  \mathcal{A}\left(H \rightarrow h' l_{\alpha} l_{\beta} h''\right)\Bigr|_{N\text{ long-lived}}
  \propto \sum_{s=\pm \frac{1}{2}} \mathcal{A}\left(H \rightarrow h' l_{\alpha} N_s(q)\right)
  \mathcal{A}\left(N_s(q) \rightarrow l_{\beta} h''\right)
\end{equation}
where we have omitted the complex phase $e^{-iq\cdot(x_D-x_P)}$ resulting from the HNL propagation,
which is unimportant in the case of one HNL.
The sub-amplitudes for the primary and secondary polarized decays are then straightforward to compute
using the usual Feynman rules with two-component spinors~\cite{dreiner_two-component_2010}.

\begin{figure}
  \begin{minipage}[t]{0.48\linewidth}
    \centering
    \includegraphics[width=\linewidth]{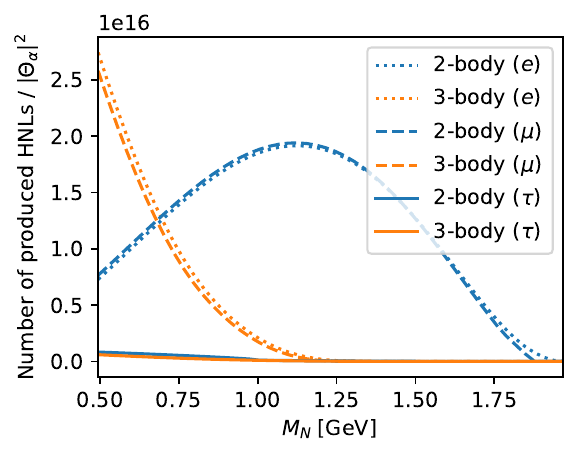}
    \caption{Number of HNLs produced at \ship{} as a function of the primary decay multiplicity, for a coupling to one flavor.}
    \label{fig:hnls_per_multiplicity}
  \end{minipage}
  \hfill
  \begin{minipage}[t]{0.48\linewidth}
    \centering
    \includegraphics[width=\linewidth]{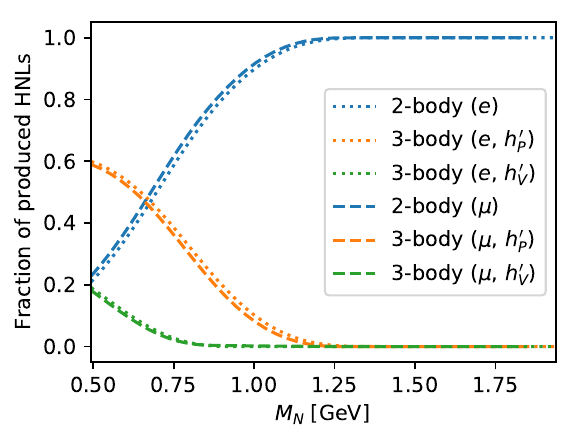}
    \caption{Fraction of HNLs produced at \ship{} as a function of the primary decay multiplicity and spin of the outgoing meson, for a coupling to one flavor.}
    \label{fig:hnl_fraction_per_multiplicity}
  \end{minipage}
\end{figure}

Consider now the LNC and LNV processes $H \rightarrow [h'] l_{\alpha} (N \rightarrow l_{\beta} h'')$
where $H, h', h''$ are pseudoscalar mesons and $h'$ may be missing.
They are respectively represented in \cref{fig:lnc_chain,fig:lnv_chain},
with the arrows denoting the flow of lepton number.
Their $CP$-conjugates have been omitted, since in the absence of oscillations
(as is the case for the incoherent width), $CP$ is conserved.
As can be seen in \cref{fig:hnl_fraction_per_multiplicity},
the primary decays $H \to [h'] l_{\alpha} N$ with $h'$ a pseudoscalar meson or missing
indeed produce the majority of HNLs with masses $\gtrsim\SI{0.7}{GeV}$ and below the $D_s$ mass.%
\footnote{%
  Below $M_N \approx \SI{0.7}{GeV}$, a non-negligible fraction of HNLs is produced along with a \emph{vector} meson.
  In this case, we expect the angular correlations to reverse compared to the pseudoscalar case.
}
Let $J_{W \mu}^h$ be the hadronic charge-lowering current,
$j_{1\mu}^- = \smash{\braket[big]{h'|J_{W \mu}^h|H}}$ and $j_{2\mu}^{\mp} = \smash{\braket[big]{h''|J_{W \mu}^{h(\dagger)}|0}}$
the hadronic matrix elements,
$p_{\alpha,\beta}$ the charged lepton momenta, and $q$ the HNL momentum.
If the primary decay is purely leptonic, then $\ket{h'} = \ket{0}$.
Since \ship{} cannot directly measure the spin or helicity of the particles detected,
we sum incoherently over all possible spin configurations of final state particles.
The spin-summed, squared amplitudes are then, in the Fermi approximation:
\begin{align}
  \overline{\big|\mathcal{A}_{\mathrm{LNC}}(H \rightarrow h' l_{\alpha}^+ l_{\beta}^- h'')\big|^2} &=
  \frac{\abs{\Theta_{\alpha}}^2 \abs{\Theta_{\beta}}^2}{v^8} \,\mathrm{tr}\left(P_R \slashed{p}_{\alpha} \slashed{j}_1^* \slashed{q} \slashed{j}_2^* \slashed{p}_{\beta} \slashed{j}_2 \slashed{q} \slashed{j}_1 \right)
  \label{eq:lnc_amplitude} \\
  \overline{\big|\mathcal{A}_{\mathrm{LNV}}(H \rightarrow h' l_{\alpha}^+ l_{\beta}^+ h'')\big|^2} &=
  \frac{\abs{\Theta_{\alpha}}^2 \abs{\Theta_{\beta}}^2}{v^8} M_N^2 \,\mathrm{tr}\left(P_R \slashed{p}_{\alpha} \slashed{j}_1^* \slashed{j}_2^* \slashed{p}_{\beta} \slashed{j}_2 \slashed{j}_1 \right)
  \label{eq:lnv_amplitude}
\end{align}
where we have omitted the ${}^\pm$ for brevity if they can be inferred from context,
$\Theta_{\alpha,\beta}$ are the mixing angles,
and $v = \vev{} \approx \SI{246}{GeV}$ is the vacuum expectation value of the Higgs field.
These results are consistent with the polarized decay rates from ref.~\cite{ballett_heavy_2019},
but generalize to the case where the primary decay produces a superposition of HNL helicity eigenstates.
The above two expressions differ in the trace,
therefore we generically expect them to produce different momentum distributions for LNC and LNV processes.
However, in their current form, this difference is not manifest.
To understand it, it is interesting to consider the special case where the production process is a two-body decay.
As can be seen in \cref{fig:hnls_per_multiplicity,fig:hnl_fraction_per_multiplicity},
it is actually the main production channel for HNLs with masses $\gtrsim\SI{1}{GeV}$ and below the $D_s$ mass.

When both the production and decay process are two-body decays, the hadronic matrix elements are
$j_1^{\mu} = -i V_{UD} f_H p_H^{\mu}$ and $j_2^{\mu} = +i V_{U'D'} f_{h''} p_{h''}^{\mu}$,
where $V_{UD}$ denotes the relevant CKM matrix element and $f_h$ is the meson decay constant.
Neglecting the masses of the final state particles,
which give $\mathcal{O}\left(\nicefrac{m_{\alpha,\beta,h''}^2}{M_{H,N}^2}\right)$ corrections,
the traces from \cref{eq:lnc_amplitude,eq:lnv_amplitude}, respectively for LNC and LNV processes, simplify to:
\begin{align}
  \mathrm{tr}\left(P_R \slashed{p}_{\alpha} \slashed{j}_1^* \slashed{q} \slashed{j}_2^* \slashed{p}_{\beta} \slashed{j}_2 \slashed{q} \slashed{j}_1 \right)
  &\cong \abs{V_{UD}}^2 \abs{V_{U'D'}}^2 f_H^2 f_{h''}^2 \cdot M_N^6 \left( M_H^2 - M_N^2 - s_{ll} \right) \\
  M_N^2 \mathrm{tr}\left(P_R \slashed{p}_{\alpha} \slashed{j}_1^* \slashed{j}_2^* \slashed{p}_{\beta} \slashed{j}_2 \slashed{j}_1 \right)
  &\cong \abs{V_{UD}}^2 \abs{V_{U'D'}}^2 f_H^2 f_{h''}^2 \cdot M_N^6 s_{ll}
\end{align}
where $s_{ll} \eqdef (p_{\alpha} + p_{\beta})^2$ is the invariant dilepton mass.
Note the linear and opposite dependences of the LNC and LNV spin-summed squared amplitudes on $s_{ll}$.
To understand their origin, it is enlightening to reexpress $s_{ll}$ in the rest frame of the HNL,
in terms of the angle $\theta_{ll}^{\mathrm{CM}} = \angle(\vec{p}_{\alpha}^{\mathrm{CM}},\vec{p}_{\beta}^{\mathrm{CM}})$
between the two lepton momenta.
Still in the massless limit, we find:
\begin{equation}
  s_{ll} = \frac{M_H^2-M_N^2}{2} \Big(1 - \cos\big(\theta_{ll}^{\mathrm{CM}}\big) \Big)
\end{equation}
Therefore,
\begin{align}
  \overline{\big|\mathcal{A}_{\mathrm{LNC}}\big|^2} &\propto 1 + \cos\big(\theta_{ll}^{\mathrm{CM}}\big)
  \label{eq:lnc_amplitude_cm} \\
  \overline{\big|\mathcal{A}_{\mathrm{LNV}}\big|^2} &\propto 1 - \cos\big(\theta_{ll}^{\mathrm{CM}}\big)
  \label{eq:lnv_amplitude_cm}
\end{align}
We observe that opposite-sign leptons (LNC) tend to be produced in the same direction,
and same-sign leptons (LNV) in opposite directions.
As explained in \cref{fig:sketch},
this is a consequence of the chirality of the weak interaction and the conservation of the total angular momentum.
In the absence of any other dynamics, spin projections lead to the characteristic angular dependence
in $\cos\left(\nicefrac{\theta_{ll}^{\mathrm{CM}}}{2}\right)$ and $\sin\left(\nicefrac{\theta_{ll}^{\mathrm{CM}}}{2}\right)$
of the transition amplitude, respectively for LNC and LNV.
\Cref{eq:lnc_amplitude_cm,eq:lnv_amplitude_cm} then directly follow from squaring the amplitude.

In the massive case, the finite masses of the decay products can result in helicity flips,
and in the three-body case, the QCD matrix elements lead to non-trivial correlations
between the momenta of the primary decay products.
These effects complicate the correlations between the various momenta.
Nevertheless, they can be accounted for when \emph{sampling} events.
To this end, we have implemented the full matrix elements from \cref{eq:lnc_amplitude,eq:lnv_amplitude}
in our Monte-Carlo simulation, as discussed in \cref{sec:spin_correlated_decay_chain}.

\subsection{Angular distribution in the laboratory frame}
\label{sec:lab_frame_distribution}

At \ship{}, the invariant mass~$s_{ll}$ (or angle~$\theta_{ll}^{\mathrm{CM}}$) cannot be reconstructed.
This is because neither the heavy hadron momentum nor the momenta of its decay products
(other than the HNL) can be determined.
Indeed, the heavy hadrons producing the HNLs do not have a monochromatic spectrum,
and the primary decay cannot be observed since it takes place inside the target.
One can then reasonably wonder if some difference between the LNC and LNV distributions subsists
when looking only at the (observable) secondary decay products, in the laboratory frame,
or if it is washed out.

\begin{figure}[h]
  \centering
  \includegraphics[width=\textwidth]{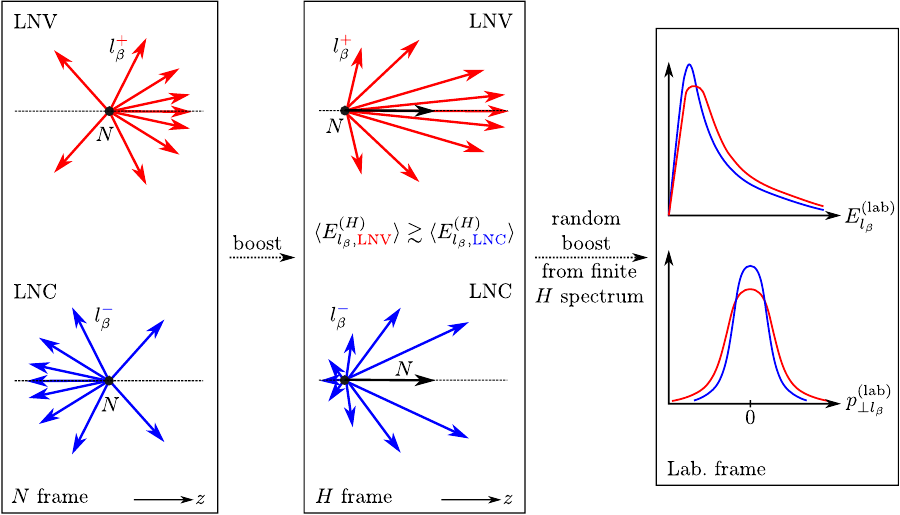}
  \caption{%
    This sketch shows how the different distributions of $l_{\beta}$ in the HNL rest frame for LNC vs.\@ LNV processes
    affect the corresponding distributions in the rest frame of the heavy hadron~$H$ and in the laboratory frame.
    The various momenta shown for $l_{\beta}$ represent multiple realizations of the decay.
    In the $H$ frame, LNV processes typically result in larger momenta for $l_{\beta}$ than LNC ones.
    In the laboratory frame, this effect partly survives the averaging over the heavy hadron spectrum
    and manifests itself as a broadening of the distribution of the secondary lepton momentum $p_{\beta}$.
  }
  \label{fig:frames}
\end{figure}

To start answering this question, it is instructive to go back to the simplified case discussed in \cref{sec:kinematics},
where the HNL is produced and decays through two-body processes involving pseudoscalar mesons.
In the HNL rest frame, we obtained the following correlation:
for LNV processes, the direction of the secondary lepton momentum is positively correlated
with the boost direction (denoted by $z$ on \cref{fig:sketch,fig:frames}) from the heavy meson rest frame to the HNL rest frame;
while for LNC processes it is anti-correlated. This is depicted in the left panel of \cref{fig:frames}.
Furthermore, in two-body decays, the magnitudes of all momenta in the rest frame of the parent particle
are fixed by four-momentum conservation, and depend only on the particle masses.
Consequently, in the heavy meson rest frame, the magnitude of the secondary lepton momentum
will on average be larger for LNV processes compared to LNC ones.
This argument is still valid for three-body decays involving pseudoscalar mesons.
A non-trivial asymmetry thus subsists in the heavy meson rest frame
(see the middle panel of \cref{fig:frames}).

As a final step, the momenta must be boosted back to the laboratory frame.
Since the heavy hadron momentum is not fixed, this has the potential to wash out the correlations.
At \ship{}, heavy mesons have a large momentum spread along the beam axis
($\mathcal{O}(\SI{10}{GeV})$, much larger than the yield of the meson decay),
and a significantly smaller one ($\mathcal{O}(\SI{1}{GeV})$) in the transverse direction
(see \cref{sec:heavy_meson_spectrum}).
The asymmetry between the LNC and LNV distributions is therefore more likely to be visible
in the transverse plane than along the beam axis.
For it to be significant, the HNL kinetic energy in the heavy hadron rest frame
should be similar to or exceed the transverse momentum spread of the hadron spectrum.
As a result, we expect the $p_T$ spectrum of the secondary lepton~$l_{\beta}$
to be broader for LNV processes than for LNC ones (see the right panel of \cref{fig:frames}),
provided that both of them are broader than the irreducible $p_T$ spread of the heavy meson spectrum. 

Alternatively, one could try to approximate the angle $\theta_{ll}^{\mathrm{CM}}$ in the HNL rest frame.
If the heavy hadron momentum is fixed,
this can be done exactly, and results in the maximal classification accuracy allowed by spin projections
(\eg{} $a=3/4$ in the two-body, massless case).
It is then equivalent to measuring the (observable) momentum $p^{\mathrm{CM}}$
of the secondary lepton $l_{\beta}$ in the HNL rest frame.
However, when the heavy hadron has a finite spectrum,
the boost direction from its rest frame to the HNL rest frame is not fixed any more.
This partially decorrelates $\theta_{ll}^{\mathrm{CM}}$ and $p^{\mathrm{CM}}$,
hence reducing the discriminating power of the latter.

As we shall see in \cref{sec:classification_summary},
the features discussed above can indeed be used to discriminate between LNC and LNV processes
(see for example \cref{fig:corrplot}).
More generally, any difference --- in the laboratory frame --- between the distributions
of the visible decay products of LNC and LNV processes
opens up the possibility of measuring their relative rates, given sufficiently many events.
Although discriminating between these two classes of events would be very challenging analytically,
this problem is well suited to multivariate analysis.

Further complications arise, however, due to HNLs being produced from a mix of various two- and three-body decays,
and because of the geometrical acceptance of the experiment, which alters the distribution of visible particles.
Generating a training set which faithfully reproduces the angular correlations discussed above
while including these effects is therefore best done using a Monte-Carlo simulation.
In the next section, we discuss the simulation used to generate the training set (\cref{sec:simulation_summary}),
then how we use it to train a binary classifier (\cref{sec:classification_summary}),
and finally how we use the classifier output in order to perform model selection (\cref{sec:model_selection_summary})
and reconstruct HNL oscillations (\cref{sec:hnl_oscillations_summary}).
In \cref{sec:other_experiments}, we discuss the applicability of the method presented here to other proposed experiments.

\section{Simulation and analysis}
\label{sec:simulation_and_analysis}

\subsection{Simulation}
\label{sec:simulation_summary}

In order to accurately estimate the distribution of the momenta of the HNL decay products,
we have devised a simple Monte-Carlo simulation,
which generates the primary and secondary decays at once,
using the matrix elements presented in \cref{sec:kinematics}.
The first step is to generate $D$~mesons with a realistic spectrum.
Generating these spectra from simulation would be a difficult undertaking,
so instead we chose to use experimental data collected by the LEBC-EHS collaboration~\cite{lebc_d-meson_1987},
at the CERN SPS running at \SI{400}{GeV} with a hydrogen target.
We then randomly select a production and decay channel
according to the relative abundances of charmed mesons from ref.~\cite{alekhin_facility_2016}
and the branching fractions from ref.~\cite{bondarenko_phenomenology_2018}.
Finally, we generate the momenta of both the primary and secondary decay products at once.
This is done by first sampling all the momenta according to phase-space, independently for each decay,
and finally performing rejection sampling on these momenta using the matrix element for the combined process.
As a last step, we simulate the geometrical acceptance by requiring
the HNL to decay within the hidden sector decay vessel,
into two long-lived, charged particles which both intersect the tracking station.
In order to account for the (small) probability of the HNL decaying inside the fiducial volume,
each event is weighted by $P_{\mathrm{decay}}(\tau) = \Gamma e^{-\Gamma\tau}$,
where $\tau$ is the proper time between the HNL production and decay.
Throughout this paper, we assume the particle identification to be perfectly efficient,
which should be a reasonably good approximation at \ship{}~\cite{hosseini_particle_2017}.
The simulation is described in details in \cref{sec:simulation}.

\subsection{LNC / LNV classification}
\label{sec:classification_summary}

For a given choice of relative squared mixing angles~$\abs{\Theta_{\alpha}}^2$
(which are supposed to be known by the time LNV is studied at \ship{}),
we generate a dataset for a range of HNL masses between the $K$ and $D_s$ thresholds.
For each HNL mass, we sample $3\cdot 10^6$~events with uniform weights,
and keep only those passing the acceptance cuts.
The HNL is simulated as a single Majorana particle,
which ensures that the dataset contains equal numbers of LNC and LNV events,
and is also balanced with respect to the primary and secondary lepton charges.

\begin{figure}
  \centering
  \includegraphics[width=\textwidth]{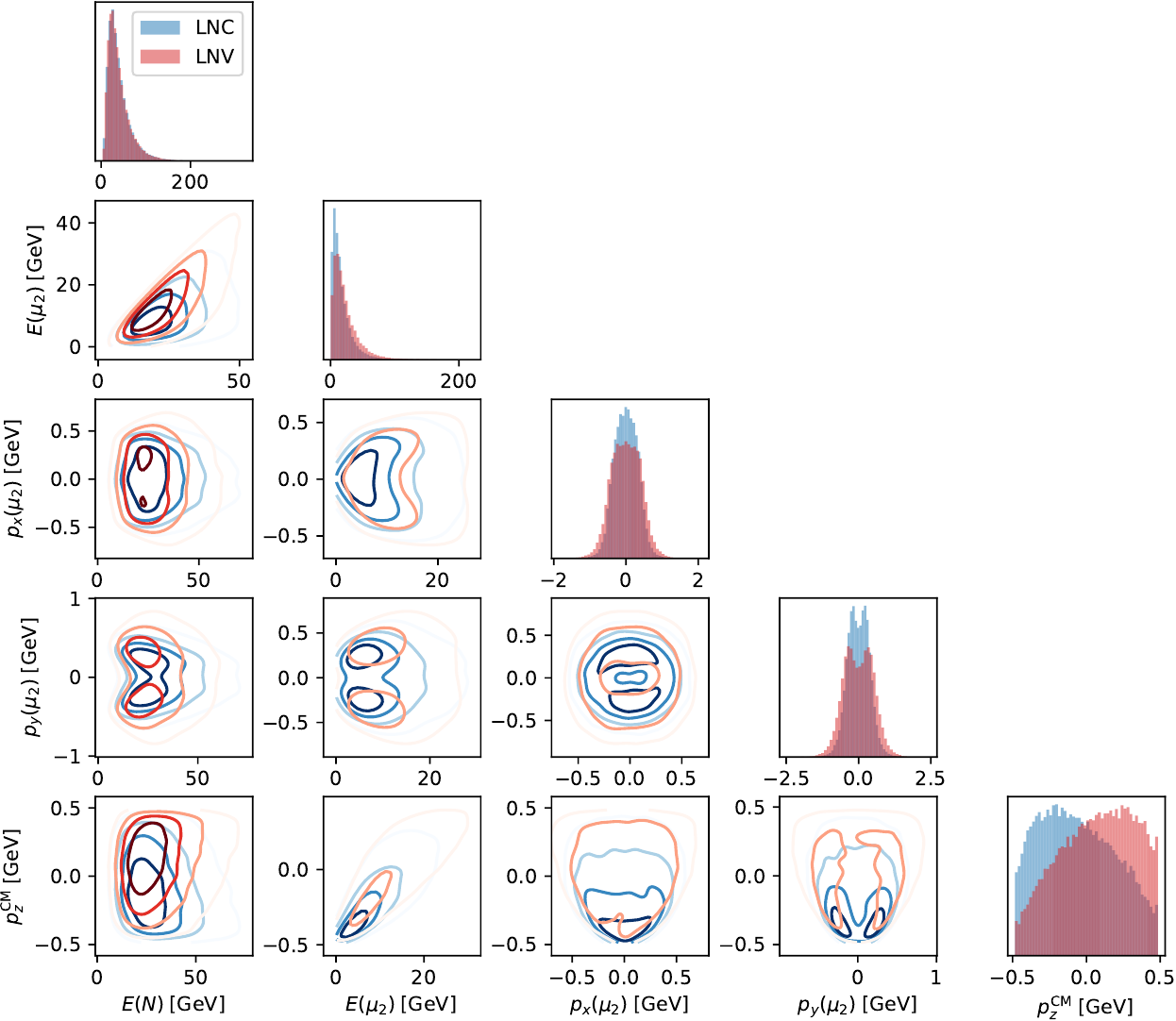}
  \caption{%
    Corner plot showing the correlations between five selected features,
    for a \SI{1}{GeV} HNL coupling to the muon.
    See \cref{tab:features} for a description of the features.
    Each subplot shows, on the same scale, the marginal distributions of LNC and LNV events
    as a function of either one (on-diagonal plots) or two (off-diagonal plots) features.
    $1\mathrm{d}$ distributions are represented as histograms,
    and $2\mathrm{d}$ distributions as contour plots of the probability density.
  }
  \label{fig:corrplot}
\end{figure}

Each event is labelled with a boolean flag set to false for LNC and true for LNV, using the MC truth.
The only observable quantities come from the HNL decay in the vacuum vessel.
They are: the momenta and charges of the lepton~$l_{\beta}^{\pm}$ and pion~$\pi^{\mp}$, and the decay vertex~$x_D$.
Of these quantities, we record a total of $19$~primary or derived features.
Their definitions can be found in \cref{tab:features},
and some typical distributions are presented, as an example, in \cref{fig:corrplot},
for both LNC and LNV processes.
Finally, from each dataset, we set aside $30\%$ of events for testing and $20\%$ for validation,
leaving us with $50\%$ of events for training the classifier.

\begin{table}
  \centering
  \begin{tabular}{|l|l|}
    \hline
    Feature(s) & Description \\
    \hline
    \texttt{Ql2} & Charge of the secondary lepton $l_{\beta}$ \\
    \texttt{E1}, \texttt{p1x}, \texttt{p1y}, \texttt{p1z} & Reconstructed HNL momentum $p_N = p_{l_{\beta}} + p_{\pi}$ (lab frame) \\
    \texttt{E2}, \texttt{p2x}, \texttt{p2y}, \texttt{p2z} & Secondary lepton momentum $p_{l_{\beta}}$ (lab frame) \\
    \texttt{E3}, \texttt{p3x}, \texttt{p3y}, \texttt{p3z} & Secondary pion momentum $p_{\pi}$ (lab frame) \\
    \texttt{pCMx}, \texttt{pCMy}, \texttt{pCMz} & Secondary lepton momentum $p_{\mathrm{CM}}$ (HNL frame) \\
    \texttt{xD}, \texttt{yD}, \texttt{zD} & Decay vertex (lab frame) \\
    \hline
  \end{tabular}
  \caption{The $19$~features recorded for each event.}
  \label{tab:features}
\end{table}

\begin{figure}
  \centering
  \begin{minipage}[t]{0.48\textwidth}
    \includegraphics[width=\linewidth]{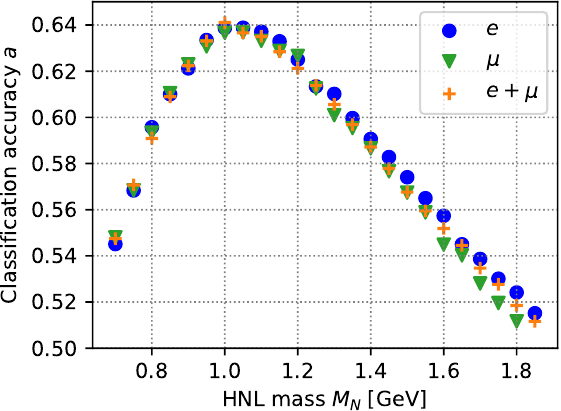}
    \caption{Classification accuracy as a function of the mass, for an HNL coupling to $e$, $\mu$, or equally to both.}
    \label{fig:classification_accuracy}
  \end{minipage}
  \hfill
  \begin{minipage}[t]{0.48\textwidth}
    \includegraphics[width=\linewidth]{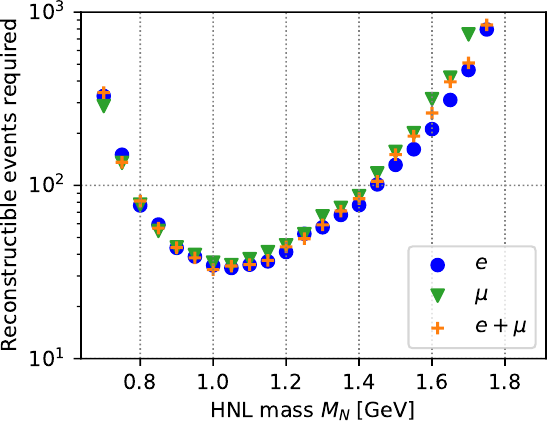}
    \caption{Number of fully reconstructible events required to detect LNV at $90\%~\mathrm{CL}$, for an HNL coupling to $e$, $\mu$, or equally to both.}
    \label{fig:required_events}
  \end{minipage}
\end{figure}

For each dataset, we train a binary classifier to discriminate between LNC and LNV decay chains.
For this study, we use the LightGBM~\cite{ke_lightgbm_2017} decision tree boosting algorithm,
through the Python interface to the reference implementation~\cite{lightgbm}.
In order to perform simple classification, we choose the \texttt{binary} objective.
The training is discussed in more details in \cref{sec:classifier}.
The accuracy of the trained classifier (as evaluated on the test set)
is presented in \cref{fig:classification_accuracy} as a function of the HNL mass for three scenarios,
corresponding to an HNL coupling to electrons, muons, or equally to both.

\subsection{Model selection}
\label{sec:model_selection_summary}

Assuming the true event distribution to match (or be sufficiently close to) the simulated one,
we can then use our trained classifier to classify each event as either LNC or LNV.
As stated in \cref{sec:introduction}, our main goal is to distinguish the following two hypotheses:
\begin{itemize}
\item $\mathcal{H}_1$: HNLs are Dirac or quasi-Dirac with $\delta M \tau \ll 1$ (LNC decays only).
\item $\mathcal{H}_2$: HNLs are Majorana or quasi-Dirac with $\delta M \tau \gg 1$ (as many LNC / LNV decays).
\end{itemize}
Since the classifier is not perfectly accurate,
its decision cannot be used to directly confirm the presence of LNV processes,
or constrain their existence.
If we knew the full distribution in feature space $\rho(z)$ for each hypothesis,
we could obtain an optimal test statistics by constructing the corresponding likelihood ratio~\cite{neyman_1933}.
However, accurately estimating~$\rho(z)$ is a non-trivial task and would be error-prone,
so we elected to use a less powerful but more reliable, simplified model.
Knowing the classification accuracy~$a$ for a given binary classifier,
we compute the likelihood of classifying $k$~events out of~$N$ as LNV, and $N-k$~events as LNC
(independently of their specific feature vectors~$z$)
assuming that the true fraction of LNV events is~$f$.
We then compute the best-fit value for~$f$ and use Wilk's theorem~\cite{wilks_large-sample_1938}
in order to determine whether it significantly deviates from either
$f=0$ (corresponding to $\mathcal{H}_1$) or $f=\frac{1}{2}$ (corresponding to $\mathcal{H}_2$).

In order to estimate the ``model-selection'' sensitivity of \ship{},
we then compute, under each hypothesis and as a function of
the HNL mass~$M_N$ and squared mixing angles~$\abs{\Theta_{\alpha}}^2$,
the median confidence level at which we can exclude the other hypothesis
assuming $5$~years of nominal operation (\ie{} $2\cdot 10^{20}$~protons on target).
For each true hypothesis, we finally draw the sensitivity limit by plotting, for each~$M_N$,
the smallest~$\abs{\Theta_{\alpha}}^2$ for which this median confidence level is at least~$0.9$.
In other words, for mixing angles above this limit, \ship{} has a probability of at least~$1/2$
of disfavouring one hypothesis at $\mathrm{CL}=0.9$ if the other is realized.
The number of fully reconstructible events corresponding to this limit is plotted in \cref{fig:required_events}
(when the null hypothesis is taken to be $\mathcal{H}_1$).
The construction of these confidence limits is described in details in \cref{sec:model_selection},
and the resulting sensitivity plots are presented in \cref{sec:results_lnv}.

\subsection{Resolving HNL oscillations}
\label{sec:hnl_oscillations_summary}

So far we have only considered the two extreme cases ($\mathcal{H}_1$ and $\mathcal{H}_2$),
where the HNL(s) behave either as a single Dirac or Majorana particle.
However, as discussed in \cref{sec:oscillations},
if two nearly degenerate HNLs form a quasi-Dirac pair,
both LNC and LNV decay chains will be present, with a non-trivial ratio $\neq 0, 1$,
and the corresponding decay rates will feature oscillations as a function of the proper time~$\tau$
between the HNL production and decay events,
with the characteristic $1 \pm \cos(\delta M \tau)$ dependence described by \cref{eq:quasi_dirac},
where ($+$) corresponds to LNC and ($-$) to LNV.

For $\delta M \sim 10^{-6}\;\si{eV}$, $\delta M \tau$ will be of order~$2\pi$ at \ship{},
leading to potentially resolvable oscillations,
provided we can accurately reconstruct the proper time~$\tau$ between the HNL production and decay.
Expressing it as $\tau = \nicefrac{L}{\beta\gamma}$,
we see that this can be accomplished if we have sufficiently accurate vertexing and energy reconstruction.
At \ship{}, the precision on~$L$ will be limited by the impossibility of reconstructing
the primary vertex within the target.
The energy resolution, despite being sufficient for particle identification,
is not enough for reconstructing~$\tau$ (see sections 4.7 and 4.10 in ref.~\cite{ship_collaboration_facility_2015}).
However, the momentum resolution, combined with the dispersion relation
(assuming the HNL mass to be known already with sufficient accuracy)
should allow reconstructing~$\gamma$ much more precisely.
The high vertexing and momentum resolution permitted by the \ship{} tracker,
together with our method for (statistically) distinguishing LNC from LNV processes
(described in \cref{sec:model_selection_summary}),
should therefore make it possible to resolve the oscillation pattern in part of the parameter space.

In order to search for HNL oscillations,
we first classify the observed events using a model trained (for \emph{one} HNL) at the corresponding mass.
We thus assume again that we have sufficiently many events that the HNL mass~$M_N$ is well known.
The events are then binned in proper time~$\tau$,
which is the relevant variable for oscillations of massive, relativistic particles.
Instead of using the predicted class,
here we implement the classifier decision as a weight for the binned events,
using the predicted probability $p_{\mathrm{LNV}}$.
This weight contains more information than the class does, since it acts as a measure of uncertainty
by taking values close to~$1/2$ for ambiguous events, and closer to~$0$ or~$1$ for unambiguous ones.
However, without applying further corrections,
the sum of these probabilities would average to~$N\braket{p_{\mathrm{LNV}}}$ for the entire sample of $N$~events.
If used directly as weights,
they would therefore cause the oscillatory pattern to be hidden among Poisson fluctuations.
In order to reveal this pattern, we instead weight the events by $p_{\mathrm{LNV}} - \overline{p_{\mathrm{LNV}}}$,
where $\overline{p_{\mathrm{LNV}}}$ is the sample average of the estimated~$p_{\mathrm{LNV}}$.
This weight averages to zero over the entire sample,
which limits the impact of Poisson fluctuations.

HNL oscillations are implemented in our simulation
by first generating events without taking interference into account then,
in a second time, performing rejection sampling based on the proper time~$\tau$,
following~\cref{eq:quasi_dirac}.
The results obtained using this simulated data set are presented in \cref{sec:results_oscillations}.

\subsection{Applicability of the method to other experiments}
\label{sec:other_experiments}

The present study crucially relies on the identification of the HNL decay products and the measurement of their momenta.
However, a number of proposed experiments to search for HNLs, such as
MATHUSLA~\cite{chou_new_2017,curtin_long-lived_2018,Alpigiani:2018fgd},
CODEX-b~\cite{gligorov_searching_2018,Aielli:2019ivi} (in its baseline configuration)
and ANUBIS~\cite{Bauer:2019vqk}, cannot measure the momenta of the decay products.
Since low-mass HNLs ($M_N < M_{B_c}$) at the LHC are also mostly produced in the decays of heavy mesons,
one can wonder to which extent the present analysis would apply to these experiments.
Training a classifier using only the \emph{directions} of the tracks of the visible decay products
and the same geometry as \ship{} reveals that the distributions of LNC / LNV for a given set of HNL parameters
can still be distinguished, with an accuracy only slightly lower than the one obtained using the full momenta.
There are, however, two caveats.
First, training the classifier requires knowing the HNL mass,
which cannot be obtained without measuring the momenta of its decay products
(or matching the displaced decay to its reconstructed production process in the main detector, if this is feasible).
In addition, the large center-of-mass energy at the LHC could result is a very broad heavy meson spectrum,
which would smear out the LNC / LNV distributions and make them indistinguishable.
It therefore seems unlikely that MATHUSLA, CODEX-b or ANUBIS could benefit from this method.

Other planned or proposed detectors, such as NA62\textsuperscript{++}~\cite{NA62:2017rwk,drewes_na62_2018} (in beam-dump mode),
the DUNE near detector~\cite{Akiri:2011dv,Krasnov:2019kdc,ballett_heavy_2019},
FASER~\cite{faser_2018,Kling:2018wct,Ariga:2018zuc} and AL3X~\cite{gligorov_leveraging_2018,Dercks:2018wum},
are in principle capable of reconstructing the HNL mass.
The AL3X detector, thanks to its large time projection chamber and its magnetic field,
should be able to directly measure both the charges and momenta of the two leptons,
making the method described here unnecessary.
It is unclear to the authors, however, whether FASER could benefit from it.
The answer likely depends on the spectrum of the heavy mesons producing the HNLs
which eventually interact with the detector.
A Monte-Carlo simulation would provide a definitive answer to this question.
The remaining beam-dump experiments: NA62\textsuperscript{++} and DUNE, share a similar geometry with \ship{}
and face the same challenge (no observation of the primary charged lepton~$l_{\alpha}^{\pm}$).
As such, we generically expect the method presented here to be applicable to these experiments,
within the mass range where it is valid, and subject to the heavy meson spectrum being similar to the one at \ship{}.
This could be ascertained using a Monte-Carlo simulation.
Whether these experiments can also resolve HNL oscillations will depend on how accurately they can reconstruct the HNL momentum.

\section{Results}
\label{sec:results}

\subsection{Sensitivity to Lepton Number Violation}
\label{sec:results_lnv}

\begin{figure}
  \centering
  \begin{subfigure}[b]{0.49\linewidth}
    \centering
    \includegraphics[width=\linewidth]{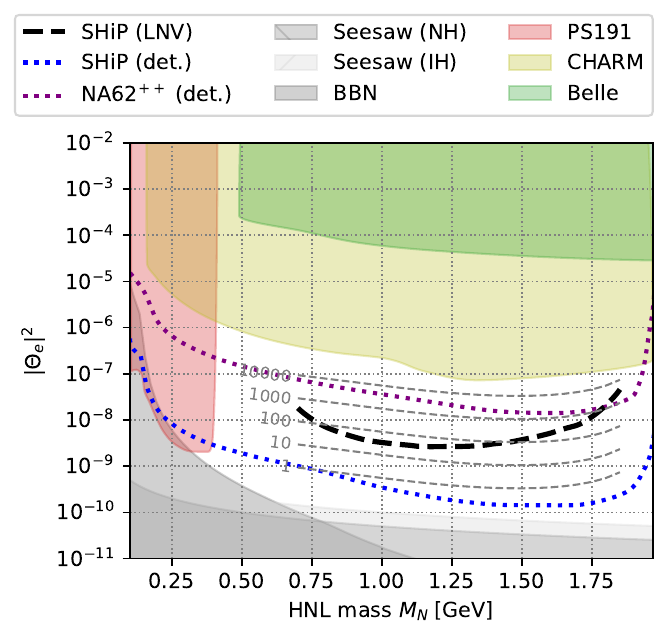}
    \caption{HNL mixing with~$\nu_e$.}
    \label{fig:lnv_sensitivity_e}
  \end{subfigure}
  \hfill
  \begin{subfigure}[b]{0.49\linewidth}
    \centering
    \includegraphics[width=\linewidth]{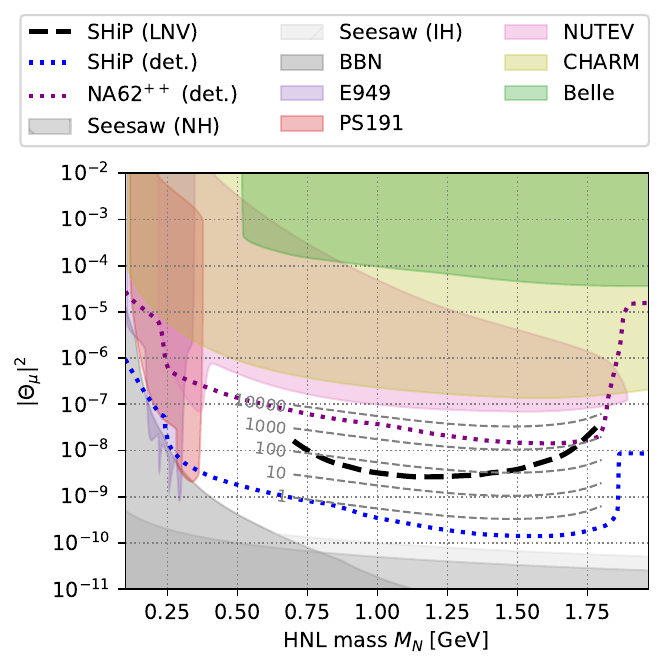}
    \caption{HNL mixing with~$\nu_{\mu}$.}
    \label{fig:lnv_sensitivity_mu}
  \end{subfigure}
  \caption[\ship{} sensitivity to lepton number violation.]{
    \ship{} sensitivity to lepton number violation.
    The thick dashed curve is the ``model-selection'' sensitivity computed in this work.
    The thin dashed grey lines show the number of fully reconstructible events which
    would be observed at SHiP for a given mass and mixing angle.
    Dotted curves are the (lower) detection sensitivities for the proposed or planned experiments
    which can reconstruct the HNL mass.
    Coloured, filled areas are regions of parameter space which have been excluded by previous experiments.
    The grey filled area denoted by BBN indicates the region which is incompatible with Big Bang Nucleosynthesis.
    Below the seesaw limit\footnotemark (hatched region),
    mixing angles are too small to produce the observed neutrino masses.
  }
  \label{fig:lnv_sensitivity}
\end{figure}

In order to easily compare our results to existing exclusion bounds or to the sensitivities of future experiments,
let us consider two simplified models
where a single HNL exclusively mixes with the electron or muon neutrino.%
\footnote{%
  Within the seesaw mechanism,
  it is impossible to generate the two observed light neutrino mass differences
  with a single HNL, or
  if HNLs mix with one generation only~\cite{drewes_minimal_2019}.
  The two benchmarks presented in \cref{fig:lnv_sensitivity_e,fig:lnv_sensitivity_mu} are thus simplifications,
  used here because they are consistent with the parametrization employed by the PBC working group.
}
As can be seen in \cref{fig:classification_accuracy},
more generic mixing patterns with the $e$ and $\mu$ flavors do not significantly degrade the classification accuracy;
therefore they should leave the limits presented below mostly unchanged.
However, if a significant fraction of HNLs is produced through mixing with the $\tau$ neutrino,
then the present analysis would need to be modified to handle secondary production of HNLs in $\tau$ decays,
including spin correlation effects.

As discussed in \cref{sec:model_selection_summary},
we define the sensitivity to lepton number violation as the smallest mixing angles
for which \ship{} has a $1/2$~probability of either rejecting or detecting LNV,
if it is respectively absent or present with the same rate as LNC.
The results are presented in \cref{fig:lnv_sensitivity},
along with various existing exclusion bounds and detection sensitivity%
\footnote{The usual sensitivity, by opposition to the sensitivity to lepton number violation discussed here.}
limits for planned or proposed experiments,
extracted from the report of the Physics Beyond Colliders working group~\cite{pbc_2019}.
We only show the sensitivities of experiments which can not only set exclusion bounds,
but also reconstruct the HNL mass, should it be observed.
Note that in order to be consistent with the \ship{} detection sensitivity, which was computed for one Majorana HNL,
we present our results for one HNL as well.
In the realistic case of $\mathcal{N} \ge 2$ HNLs,
both curves must be scaled down by a factor of~$\mathcal{N}^{1/2}$.
Above the black dashed line,
\ship{} should be able to distinguish Dirac-like ($\mathcal{H}_1$) and Majorana-like ($\mathcal{H}_2$) HNLs.
We have discarded the HNL masses for which the early stopping criterion returned the first iteration as the best,
since it suggests that the classifier has failed to learn anything about the data.
Below $\SI{0.7}{GeV}$, additional production channels $H \to h_V' l_{\alpha} N$
(where $h_V'$ denotes a vector meson) become significant,
and have not been implemented with spin correlations in our Monte-Carlo simulation.
Therefore we also restrict the HNL mass to $M_N \gtrsim \SI{0.7}{GeV}$.
Additionally, since the sensitivity is almost identical for excluding $\mathcal{H}_1$ or $\mathcal{H}_2$,
we only plot one limit, which corresponds to excluding $\mathcal{H}_1$ at $90\%~\mathrm{CL}$ if LNV is actually present.

We can see that the larger number of accepted events (indicated in \cref{fig:lnv_sensitivity} by the thin dashed grey lines) at higher masses
initially compensates for the worse classification accuracy,
but is not sufficient any more as we approach the $D$~threshold.
In practice, we expect that systematic uncertainties about the $D$~spectrum and the simulation
will decrease the sensitivity at both ends of the mass range,
where the classification accuracy is already close to~$1/2$.
Comparing the results to the \ship{} detection sensitivity,
we see that around~\SI{1}{GeV},
the model-selection sensitivity limit is about one order of magnitude above the detection one,
while remaining well below the planned NA62${}^{++}$ limit as well as existing bounds.

This leads us to an interesting conclusion:
there exists a non-trivial region of parameter space,
unconstrained by current or near-future experiments,
where \ship{} would not only be able to detect HNLs,
but also characterize them as either Dirac-like or Majorana-like particles.
As discussed in \cref{sec:heavy_meson_spectrum,sec:spectrum_systematics},
this conclusion is robust with respect to uncertainties on the heavy meson spectrum.

\footnotetext{%
  The seesaw limit can only be rigorously computed if
  the mixing angles are consistent with the seesaw equation~\eqref{eq:seesaw}.
  This is not possible for HNLs mixing with only one generation, nor for a single HNL.
  The limits presented here instead correspond to the ``naive'' estimate
  $\sum m_{\nu} \leq M_N \cdot \sum_{\alpha} |\Theta_{\alpha}|^2$,
  where we have assumed the lightest neutrino to be massless.
}

\subsection{Resolvable quasi-Dirac oscillations}
\label{sec:results_oscillations}

\begin{figure}
  \centering
  \includegraphics[width=0.7\textwidth]{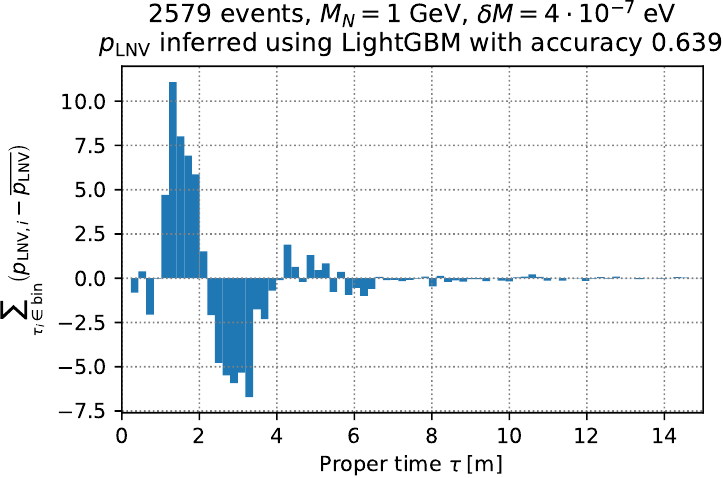}
  \caption{Events binned by proper time~$\tau$ and weighted by $p_{\mathrm{LNV}}-\overline{p_{\mathrm{LNV}}}$,
    revealing the oscillatory pattern, for \emph{two} HNLs with
    $M_N = \SI{1}{GeV}$, $\abs{\Theta_{\mu I}}^2 = 2\cdot 10^{-8}$,
    $\abs{\Theta_{e I}}^2 = \abs{\Theta_{\tau I}}^2 = 0$ and $\delta M = \SI{4e-7}{eV}$.}
  \label{fig:reconstructed_oscillations}
\end{figure}

The result of the procedure described in \cref{sec:hnl_oscillations_summary}
is presented in \cref{fig:reconstructed_oscillations}
for a new simulated dataset (independent from the training set),
corresponding to a quasi-Dirac pair of mass $M_N = \SI{1}{GeV}$, mass splitting $\delta M = \SI{4e-7}{eV}$,
and mixing with muon neutrinos only, with a squared mixing angle $\abs{\Theta_{\mu I}}^2 = 2\cdot 10^{-8}$, $I=1,2$.
The oscillatory pattern is manifest at $\tau < \SI{5}{m}$, where most of the events fall.
At larger~$\tau$ it is hidden in Poisson fluctuations.
The uncertainty on $\tau$ at \ship{} is dominated by the (boosted) length of the target $\sim\SI{0.1}{m}$,
which contains the unresolved primary vertex.
It could smear out fast oscillations,
in which case an accurate treatment of this uncertainty would be needed in the simulation.
However, for longer oscillation periods like the one shown in \cref{fig:reconstructed_oscillations},
its effect should be negligible.
Deriving precise sensitivity limits for HNL oscillations is beyond the scope of this paper,
since it is likely that no simple analytical expression exists for them,
due to the more complex test statistics required, compared to the detection or model-selection limits.
HNL oscillations might for instance be amenable to methods such as
maximum likelihood estimation, wavelets, or matched filtering,
for which the null distribution can be estimated numerically
using a (computationally expensive) bootstrapping procedure.

\section{Conclusions}
\label{sec:conclusions}

The \ship{} experiment is set to have an unprecedented detection reach
for a variety of models containing feebly interacting particles, such as Heavy Neutral Leptons (HNLs).
A distinctive feature of \ship{} among other intensity frontier experiments is its decay spectrometer,
which allows it to not only place exclusion bounds, but also perform event reconstruction and measure the HNL properties.
The simplest consistent HNL model accessible at \ship{} contains two nearly degenerate HNLs,
which can undergo oscillations.
Their mass splitting $\delta M$ is of particular interest,
since it greatly influences their phenomenology as well as early-Universe cosmology
(specifically, baryogenesis and dark matter production).

In the present work, we have investigated to which extent \ship{} may be able to constrain or even measure $\delta M$.
Depending on the scale of the oscillation phase $\delta M \tau$ accessible at an experiment,
HNLs may or may not exhibit lepton number violation (LNV).
The problem thus amounts to distinguishing LNC from LNV decay chains (\cref{fig:lnc_lnv_chains}) in a beam-dump setting (\cref{fig:ship}),
where the primary lepton cannot be observed.
We have shown that the angular distribution of the \emph{visible} secondary decay products
provides a partial solution to this problem, since, depending on the HNL mass,
it can significantly differ between LNC and LNV in the laboratory frame (\cref{fig:corrplot}).
This result has been qualitatively understood in the simplified case
of two-body decays in the massless limit (\cref{fig:sketch,fig:frames}).
In order to handle more realistic cases, a Monte-Carlo simulation has been employed
to generate accurate data sets of LNC and LNV events, including spin correlations and geometrical acceptance.
The different distributions of the kinematic variables thus allow
discriminating between LNC and LNV events using multivariate analysis;
and with sufficiently many events, it becomes possible to statistically detect or exclude lepton number violation.

In order to produce sufficiently accurate training sets, our simulation must satisfy several requirements.
It should be able to generate all the relevant two- and three-body meson decays containing an HNL (\cref{fig:hnls_per_multiplicity}),
as well as the selected HNL decay channel $N \to \pi^{\mp} l_{\beta}^{\pm}$.
It should be accurate for \si{GeV}-scale HNLs,
and should account for the spin correlations between the primary and secondary decays.
Finally, it should run sufficiently fast to allow producing large training sets for various hypotheses and parameters.
In order to meet all these requirements, we have written our own Monte-Carlo simulation,
the output of which is used to train a binary classifier.

Knowing the accuracy of the classifier decision (\cref{fig:classification_accuracy})
for a given mass and (relative) mixing angles,
we can finally draw a ``model-selection'' sensitivity limit in the $(M_N, \abs{\Theta}^2)$ plane
(shown in \cref{fig:lnv_sensitivity_e,fig:lnv_sensitivity_mu}),
above which \ship{} should be able to either discover or rule out lepton number violation from HNLs.
Interestingly, this limit lies below the detection sensitivity of near-future experiments such as NA62\textsuperscript{++}.
This leads to a striking conclusion: \ship{} might be able to not only \emph{discover} HNLs,
but also \emph{characterize} them as either ``Dirac-like'' or ``Majorana-like'' fermions
(depending on whether they feature LNV) even if previous experiments \emph{see no signal at all}.
Better yet, if the mass splitting between the two HNLs is of order $\delta M \sim 10^{-6}\;\si{eV}$,
\ship{} should be able to \emph{resolve the oscillations} of HNLs (\cref{fig:reconstructed_oscillations}),
given sufficiently many events.
Intriguingly, this mass splitting falls within the range required
for producing dark matter in the \nuMSM{}~\cite{Canetti:2012kh}.
Its measurement --- or constraining --- would therefore be an important test of cosmological models.

\acknowledgments

The authors would like to thank Mikhail Shaposhnikov for his suggestion of investigating HNL oscillations at SHiP,
Oleg Ruchayskiy and Maksym Ovchynnikov for their helpful comments and proofreading of the present manuscript,
Sonia Bouchiba and Federico Leo Redi for stimulating discussions about helicity effects in HNL decays,
and Kyrylo Bondarenko and Elena Graverini for sharing with us their code to compute HNL decay widths.
We thank the \ship{} collaboration for their interest in our work.
We are grateful to Annarita Buonaura, Richard Jacobsson, and Nicola Serra for their helpful comments and clarifications about the dimensions of the facility.
JLT acknowledges support from the Carlsberg foundation. The work of IT was supported in part by the ERC-AdG-2015 grant 694896.

\appendix

\section{Simulation}
\label{sec:simulation}

\subsection{Overview}

It is not obvious whether the different angular correlations of LNC and LNV events
lead to an observable effect in a realistic beam-dump experiment.
To answer this question, we have devised a toy Monte-Carlo simulation,
inspired from the one used in ref.~\cite{ship_collaboration_sensitivity_2018},
to simulate the production and decay of HNLs at the \ship{} experiment
\cite{ship_collaboration_facility_2015,alekhin_facility_2016}
(represented on \cref{fig:ship}).

The simulation of rare BSM processes with spin correlations entails two main requirements.
First, we cannot afford to simulate all the possible processes,
since, due to the small HNL mixing angles,
the decay chains mediated by an HNL only represent a tiny fraction of all decays.
Instead, we only simulate the BSM processes,
and use importance sampling (\ie{} introduce weights) in order to obtain the correct absolute number of events
and expectation values (\cref{sec:decay_chain}).

Secondly, we cannot sample the primary and secondary decays separately,
since they are not independent.
Instead, we construct all possible decay chains for the production and decay processes of interest,
and sample the entire chain at once, with a probability proportional to its combined branching fraction.
The momenta of all the decay products are then sampled simultaneously, using the matrix element for the entire chain
(\cref{sec:spin_correlated_decay_chain}).

In addition, in order to accurately model the \ship{} experiment,
we need to sample the heavy meson momenta from a realistic spectrum (\cref{sec:heavy_meson_spectrum})
and take into account the finite size of \ship{} and its geometrical acceptance (\cref{sec:geometry}).
Finally, since most machine learning algorithms take unweighted data points as input,
it is necessary to perform a last step of rejection sampling in order to produce
a training set consisting of events with equal weights (\cref{sec:unweighting}).

\subsection{Decay chains}
\label{sec:decay_chain}

As discussed in ref.~\cite{bondarenko_phenomenology_2018},
the dominant HNL production process at \ship{} is from weak decays of the lightest charmed or beauty mesons.
In the present study, we focus on HNL masses below the $D_s$ mass,
and only select the fully reconstructible secondary decays $N \rightarrow \pi^{\pm} l_{\beta}^{\mp}$,
By producing long-lived, charged particles which can be measured by
the decay spectrometer located at the end of the decay vessel,
they allow the HNL momentum to be reconstructed.
The efficiency of particle identification at \ship{} is high enough~\cite{hosseini_particle_2017}
that we can approximate it as one for the present estimate.
Therefore we do not need to simulate decay chains containing any other secondary decays.

For the mixing angles of interest (\ie{} below existing bounds),
the fraction of all decays which are mediated by an HNL is tiny.
We therefore need to use importance sampling in order to efficiently simulate only the processes of interest.
For every proton on target (POT), the probability of producing a charmed hadron of species~$H$ is:
\begin{equation}
  P(H) = \frac{\sigma_{cc}}{\sigma_{pN}} \cdot A_H
\end{equation}
where $\sigma_{cc}$ is the production cross-section for charmed hadrons,
$\sigma_{pN}$ the interaction cross-section for protons hitting the target nuclei,
and $A_H$ is the relative abundance of the charmed hadron species~$H$
(as given in appendix A of~\cite{alekhin_facility_2016}).
The \emph{nominal} (\ie{} physical) probability of producing an HNL which mediates
a given decay chain $H \rightarrow [h'] l_{\alpha} (N \rightarrow l_{\beta} h'')$
(irrespective of whether the decay is observed in the detector) is then:
\begin{multline}
  P\left(H \rightarrow [h'] l_{\alpha} (N \rightarrow l_{\beta} h'')\right)
  = P(H) \cdot P(h' l_{\alpha} N | H) \cdot P(l_{\beta} h'' | h' l_{\alpha} N) \\
  = \frac{\sigma_{cc}}{\sigma_{pN}} \cdot A_H \cdot \mathrm{Br}_{\mathrm{prod}}(H \rightarrow [h'] l_{\alpha} N) \cdot \mathrm{Br}_{\mathrm{decay}}(N \rightarrow l_{\beta} h'')
\end{multline}
where the last two terms are the production and decay branching ratios for HNLs in the considered decay chain.
The \emph{importance} distribution~$P'$ is defined as a uniform scaling for decay chains involving an HNL,
and as zero for all other outcomes:
\begin{equation}
  \begin{cases}
    P'\left(H \rightarrow [h'] l_{\alpha} (N \rightarrow l_{\beta} h'')\right)
    = \frac{1}{w_{\mathrm{prod}}} P\left(H \rightarrow [h'] l_{\alpha} (N \rightarrow l_{\beta} h'')\right) \\
    P'(\text{no HNL}) = 0
  \end{cases}
\end{equation}
where $w_{\mathrm{prod}}$ is the weight to be applied to all the chains sampled from the importance distribution,
and corresponds to the total probability of producing an HNL according to the nominal distribution:
\begin{equation}
  w_{\mathrm{prod}} = \sum_{\mathrm{chains}} P\left(H \rightarrow [h'] l_{\alpha} (N \rightarrow l_{\beta} h'')\right)
\end{equation}

When computing expected numbers of events over the entire duration of the \ship{} experiment,
which represents an integrated $N_{\mathrm{POT}} = 2\cdot 10^{20}$ protons on target for $5$~years of nominal operation,
we must further multiply by $N_{\mathrm{POT}}$ the expectation values obtained for one event.
This is most easily done by simply multiplying the total weights by $N_{\mathrm{POT}}$.

\subsection{Heavy meson spectrum}
\label{sec:heavy_meson_spectrum}

Once a chain is selected, we sample the momentum of the corresponding charmed meson from the spectrum
measured by the LEBC-EHS collaboration \cite{lebc_d-meson_1987}
at the CERN SPS running at \SI{400}{GeV} with a hydrogen target.
The differential cross-section is parametrized as
the product of a $\beta$~distribution in~$x_F$ and an exponential distribution in~$p_T^2$:
\begin{equation}
  \frac{\diff^2 \sigma}{\diff x_F \diff p_T^2} = \sigma \frac{(n+1)b}{2} (1 - \abs{x_F})^n e^{-b p_T^2}
\end{equation}
with the best-fit values $n = 4.9 \pm 0.5$ and $b = (1.0 \pm 0.1)\;\si{GeV^{-2}}$.
We thus implicitly assume the spectrum to be separable.
Due to their very similar mass, and to compensate for the lack of data,
we assume $D_s$~mesons to share the same spectrum as $D$~mesons.

\begin{figure}
  \centering
  \includegraphics[width=0.65\linewidth]{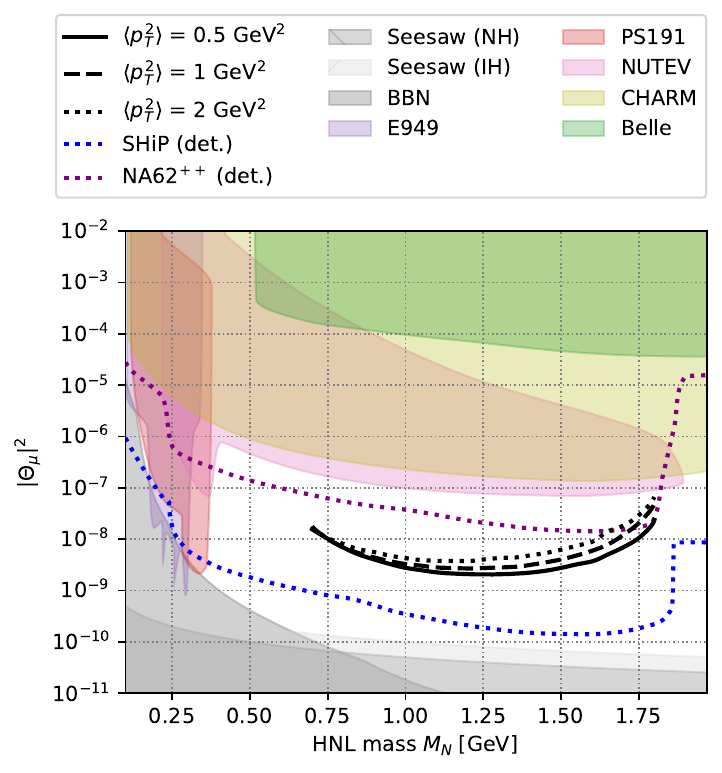}
  \caption{%
    Effect of varying the width of the heavy meson $p_T$~spectrum on the
    sensitivity to lepton number violation ($90\%$~CL), for an HNL coupling to the muon.
    Black lines represent the model-selection sensitivity of \ship{} for various values of~$\braket{p_T^2}$.
    The dashed line corresponds to the best-fit value $\braket{p_T^2} = \SI{1}{GeV^2}$ from the LEBC-EHS collaboration~\cite{lebc_d-meson_1987}.
  }
  \label{fig:pt_variation}
\end{figure}

By using the spectrum for a hydrogen target,
we effectively neglect cascade production of heavy hadrons inside the target,
leading us to underestimate the number of hadrons produced at the low-energy end of the spectrum.
This could be problematic if their $p_T$~spectrum happens to be significantly different from that of
primary hadrons produced in $pp$~collisions.
However, the lower acceptance for these softer hadrons should help mitigate the issue.
In \cref{fig:pt_variation}, we show how varying the width of the heavy meson $p_T$~spectrum
affects the final sensitivity.
As expected, a larger $p_T$~spread reduces the sensitivity, while a narrower spectrum improves it.

\subsection{Decay product momenta}
\label{sec:spin_correlated_decay_chain}

In order to preserve spin correlations between the HNL siblings and its decay products,
we simulate both the HNL production and decay processes at once.
For the masses and mixing angles of interest, the HNL is long-lived and can be assumed to be on its mass shell.
Therefore the phase-space sampling can be performed independently for the primary and secondary decays.
We use the $m$-generator algorithm~\cite{james_monte_1968} for that, as described in ref.~\cite{ilten_electroweak_2014}.
In order to sample events with a probability proportional to the squared transition amplitude,
we then perform rejection sampling, taking the phase-space distribution as proposal distribution,
and an acceptance probability proportional to the spin-summed,
squared matrix elements \eqref{eq:lnc_amplitude} and \eqref{eq:lnv_amplitude} for the entire decay chain.
Only the spin states of the external particles
(which interact with the detector and are thus ``measured'' in the quantum mechanical sense)
are summed over.

\subsection{Geometry}
\label{sec:geometry}

In order to model the geometry of the \ship{} experiment,
we must account for the finite size of the detector and its geometrical acceptance.
In the current \ship{} design
(represented on \cref{fig:ship}),
the fiducial volume consists of an evacuated right pyramidal frustum of length \SI{50}{m},
located at a distance of \SI{50}{m} from the target,
and with horizontal and vertical sides \SI{5}{m} and \SI{10}{m} respectively at the far end.
It is followed by a \SI{10}{m} long tracking station.

To estimate the probability of the HNL decaying within the fiducial volume and passing the acceptance cuts,
we use once again importance sampling for sampling the decay vertex.
This is required in order to overcome the potentially very long lifetime of HNLs,
which could cause most of them to decay away from the experiment.
We choose an importance distribution (approximately) covering the fiducial volume,
by sampling the decay vertex uniformly along the HNL momentum,
at a distance such that it falls inside the decay vessel.
The nominal decay probability density is,
as a function of the proper time~$\tau$ (or boost factor~$\gamma$ and distance~$L$)
between the HNL production and decay:
\begin{equation}
  P_{\mathrm{decay}}(\tau) = \Gamma e^{-\Gamma \tau}
  \quad\Longrightarrow\quad
  P_{\mathrm{decay}}(L | \gamma) = \frac{\Gamma}{\beta \gamma} e^{-\frac{\Gamma L}{\beta \gamma}}
\end{equation}
The partial weight resulting from this importance sampling step is therefore:
\begin{equation}
  w_{\mathrm{decay}}(L|\gamma) = \frac{\Gamma L_{\mathrm{DV}}}{\beta \gamma \cos(\theta)} e^{-\frac{\Gamma L}{\beta \gamma}}
\end{equation}
where $L_{\mathrm{DV}} = \SI{50}{m}$ is the length of the decay vessel
and $\theta$ the angle between the HNL momentum and the beam axis.
In the linear regime, where $\Gamma \tau \ll 1$, this partial weight reduces to
$w_{\mathrm{decay}}(L|\gamma) \cong \nicefrac{\Gamma L_{\mathrm{DV}}}{\beta \gamma \cos(\theta)}$.

We finally apply acceptance cuts by requiring the HNL to decay within the decay vessel,
and the trajectories of its two decay products ($l_{\beta}^{\mp}$ and $\pi^{\pm}$) to intersect
the tracking station located at its far end.

\subsection{Unweighting}
\label{sec:unweighting}

As a last step, we perform again rejection sampling on the weighted events
in order to obtain a set of events with equal weights,
which are easier to analyse and process with machine learning algorithms.
This is done by accepting events with a probability proportional to their weight,
and can be justified as follows.

Let $X$ denote a random variable representing the simulated event, and $x$ a concrete realization of it.
Let $f(x) = P(X=x)$ be the nominal (\ie{} true) distribution and $g(x)$ the importance distribution,
such that $g(x) > 0$ for all outcomes $x$ in the domain of interest~$\Omega$
(\ie{} all relevant observables must have their support in~$\Omega$).
If $x$ is sampled from the importance distribution~$g(x)$, its associated weight will be $w(x) = f(x) / g(x)$.
Let $M$ be an upper bound on $w(x)$, \ie{} $M \ge w(x), \forall x \in \Omega$.
If we choose the acceptance probability to be $a(x) \eqdef w(x) / M \leq 1$,
then it immediately follows that the accepted events,
effectively drawn from the new importance distribution $g(x) \cdot a(x)$, will have uniform weight~$M$.

It is therefore possible to perform rejection sampling a posteriori in order to produce uniformly weighted events.
However, storing all the generated events, many of which will eventually be rejected,
would be inefficient from a memory perspective.
A more economical solution, which we decided to use,
consists in performing rejection sampling directly as events are being generated.
This requires estimating an upper bound~$M$ on the weights, during an initial burn-in phase.

\section{LNC/LNV classification}
\label{sec:classification}

At leading order in the light lepton and hadron masses,
the matrix elements for LNC and LNV decay chains have a straightforward analytical dependence
on the invariant mass~$s_{ll}$ of the charged lepton pair.
However, unlike in collider experiments, this variable is not readily available in a beam-dump setting,
due to the primary lepton being unobservable.
As we saw in \cref{sec:kinematics}, the different angular correlations between the charged leptons
can nevertheless lead to residual correlations between the visible HNL decay products.
The absence of an obvious test statistics,
along with the almost background-free conditions and highly efficient PID at~\ship{} \cite{hosseini_particle_2017},
makes the task of distinguishing LNC from LNV ideally suited for multivariate analysis.
In the following subsections, we describe how we generate the training set (\cref{sec:dataset}),
the classifier used to discriminate between LNC and LNV events (\cref{sec:classifier}),
how to produce a sensitivity limit from its output (\cref{sec:model_selection}),
and finally how sensitive is the classification to systematic uncertainties
on the heavy meson spectrum (\cref{sec:spectrum_systematics}).

\subsection{Dataset}
\label{sec:dataset}

As mentioned in \cref{sec:classification_summary},
we need to generate datasets for various HNL masses~$M_N$ and rays in $\abs{\Theta_{\alpha}}^2$~space,
where $\alpha=e,\mu$ (the overall normalization does not matter).
In practice, we choose a mass range spanning the region between the $K$ and $D_s$ thresholds,
and consider several benchmark models with fixed $\abs{\Theta_e}^2:\abs{\Theta_{\mu}}^2$ ratios.%
\footnote{%
  We do not consider HNL production through $\tau$~mixing in this work,
  since it would have required to implement secondary production from $\tau$ decays.
  It is negligible in the considered mass range
  unless the $\Theta_{\tau}$ mixing angle is significantly larger than the others,
  as can be seen in \cref{fig:hnls_per_multiplicity}.
  In addition, visible HNL decays through $\tau$ mixing are forbidden below the $\tau$~threshold.
}
For each choice of physical parameters,
we sample $3\cdot 10^6$~events with uniform weights.
This is done by sampling sufficiently many weighted events and, as they are being generated,
``unweighting'' them by performing rejection sampling with an acceptance probability proportional to their weight.
Only events which pass the acceptance cuts are used for training.
In the simulation, the HNL is taken to be a single Majorana particle,
such that the dataset contains equal numbers of LNC and LNV events
and is balanced with respect to the primary and secondary lepton charges.
We select only the fully reconstructible HNL decays $N \rightarrow \pi^{\mp} l_{\beta}^{\pm}$,
which do not contain an unobservable light neutrino,
and produce long-lived charged particles which can be measured by the decay spectrometer.
For the sake of simplicity, we will assume the PID to be perfectly efficient throughout this analysis.
Non-trivial efficiencies are expected to slightly reduce the final sensitivity reach.
As explained in \cref{sec:classification_summary},
each event is labelled as being either LNC or LNV,
and we record the $19$~observable features listed in \cref{tab:features}.
The dataset is split into training / validation / test sets with respective proportions~$0.5:0.2:0.3$.

\subsection{Classifier}
\label{sec:classifier}

We employ the LightGBM~\cite{ke_lightgbm_2017} gradient boosting algorithm,
accessed through the Python interface to the reference implementation~\cite{lightgbm}.
For classification, we choose the \texttt{binary} objective.
We use early stopping based on the binary log-loss (\texttt{binary\_logloss})
and the area-under-curve (\texttt{auc}) metrics,
with a $10$~round threshold.
The hyperparameters \texttt{num\_leaves} and \texttt{learning\_rate} are manually optimized
by maximizing the above two metrics on the validation set.
The classification accuracy is presented in \cref{fig:classification_accuracy}
as a function of the HNL mass~$M_N$ for two orthogonal scenarios,
corresponding to the HNL coupling exclusively to electrons ($\abs{\Theta_e}^2:\abs{\Theta_{\mu}}^2:\abs{\Theta_{\tau}}^2=1:0:0$) or muons ($\abs{\Theta_e}^2:\abs{\Theta_{\mu}}^2:\abs{\Theta_{\tau}}^2=0:1:0$),
and a third one where it couples equally to both ($\abs{\Theta_e}^2:\abs{\Theta_{\mu}}^2:\abs{\Theta_{\tau}}^2=1:1:0$).

It is instructive to understand the origin of this dependence,
if only to make sure that it corresponds to a physical effect.
LightGBM provides a way to estimate the feature importance,
by counting the number of times a feature is used to split a tree.
Those are listed in \cref{tab:feature_importance} for a \SI{1}{GeV} HNL coupling to muons
(which results in a classification accuracy of $63.5\%$).
They reveal that the most important features are the transverse components of the momenta of the HNL decay products.
Indeed, it is possible to successfully train a model using a single feature
such as the transverse momentum $p_{T,\mu}$ of the secondary muon,
while still obtaining a classification accuracy of $61.5\%$ (for the same dataset).

Inspecting the results more closely (see \cref{fig:corrplot}) shows that
LNV events have on average a slightly larger transverse momentum than LNC ones.
This is consistent with our discussion from \cref{sec:kinematics},
and allows us to understand the mass dependence.
At large HNL masses, as we approach the closing mass of $D$~meson leptonic decays,
the kinetic energy of the HNL in the heavy meson rest frame decreases,
until it becomes so small that the difference between LNC and LNV
becomes negligible compared to the transverse momentum spread of the heavy meson spectrum.
As the HNL mass decreases, 3-body semileptonic decay channels open, and become dominant at lower masses.
The additional meson takes away part of the energy from the HNL,
leaving it with insufficient kinetic energy to ``escape'' the transverse momentum spread of the heavy meson spectrum.
Finally, the large boost of the heavy mesons along the beam axis washes out most of the information
contained in the longitudinal part of all laboratory frame momenta,
which explains their low importance.

\begin{table}
  \centering
  \begin{tabular}{|l|c|c|c|c|c|c|c|c|c|c|}
    \hline
    Feature & \texttt{p2y} & \texttt{p3y} & \texttt{p2x} & \texttt{p3x} & \texttt{pCMz} & \texttt{zD} & \texttt{xD} & \texttt{yD} & \texttt{p1x} & \texttt{pCMy} \\
    \hline
    \# splits & 302 & 282 & 243 & 238 & 141 & 114 & 105 & 97 & 91 & 85 \\
    \hline
  \end{tabular}\\
  \medskip
  \begin{tabular}{|l|c|c|c|c|c|c|c|c|c|}
    \hline
     Feature & \texttt{pCMx} & \texttt{p1y} & \texttt{E1} & \texttt{E2} & \texttt{E3} & \texttt{p3z} & \texttt{p2z} & \texttt{p1z} & \texttt{Ql2} \\
    \hline
    \# splits & 77 & 74 & 69 & 67 & 61 & 53 & 34 & 14 & 9 \\
    \hline
  \end{tabular}
  \caption{Feature importance for a \SI{1}{GeV} HNL coupling to $\mu$.}
  \label{tab:feature_importance}
\end{table}

\subsection{Sensitivity to lepton number violation}
\label{sec:model_selection}

As stated in \cref{sec:model_selection_summary},
our main goal is to distinguish between the following two hypotheses
using exclusively the classifier decision (\ie{} not the underlying feature vector~$z$):
\begin{itemize}
\item $\mathcal{H}_1$: HNLs are Dirac or quasi-Dirac with $\delta M \tau \ll 1$ (LNC decays only).
\item $\mathcal{H}_2$: HNLs are Majorana or quasi-Dirac with $\delta M \tau \gg 1$ (LNC and LNV decays).
\end{itemize}
Those can be expressed as special cases of a more general hypothesis~$\mathcal{H}(f)$, $f \in [0, 1]$,
parametrized by the relative frequency~$f$ of LNV events:
\begin{itemize}
\item $\mathcal{H}(f)$: $(\text{LNV rate}) = f \times (\text{total rate})$.
\end{itemize}
such that $\mathcal{H}_1 = \mathcal{H}(f=0)$ and $\mathcal{H}_2 = \mathcal{H}(f=1/2)$.

We model the classifier decisions using a $2\times 2$ confusion matrix
$C_{ij} = P(i \text{ classified as } j)$,
where $i,j=1,2$ correspond to the two classes, respectively LNC and LNV.
The confusion matrix can be expressed in terms of the classification accuracies as:
\begin{equation}
  C = \begin{pmatrix}
        a_1 & 1 - a_1 \\
    1 - a_2 &     a_2
    \end{pmatrix}
\end{equation}
Suppose we observe $N$~events passing the selection cuts,
$k$~of which are classified as LNV.
Then, under $\mathcal{H}(f)$,
the likelihood of classifying $N-k$~events in class~1 (LNC) and $k$ in class~2 (LNV)
is given by the following binomial distribution:
\begin{equation}
  \mathcal{L}(k;f) = {N \choose k} \big( a_2 f + (1-a_1) (1-f) \big)^k \big( a_1 (1-f) + (1-a_2) f \big)^{N-k}
\end{equation}
Under hypothesis~$\mathcal{H}_1$, \ie{} all events are LNC, this likelihood reduces to:
\begin{equation}
  \mathcal{L}_1(k) = \mathcal{L}(k;f=0) = {N \choose k} (1-a_1)^k a_1^{N-k}
\end{equation}
while under hypothesis~$\mathcal{H}_2$, \ie{} events come from either class with equal probability, it becomes:
\begin{equation}
  \mathcal{L}_2(k) = \mathcal{L}(k;f=1/2) = {N \choose k} \frac{(1+a_2-a_1)^k (1+a_1-a_2)^{N-k}}{2^N}
\end{equation}
For many models, including LightGBM (with a balanced training set), $a_1 \approx a_2 \eqdef a$.
In this limit, $\mathcal{L}_2(k)$ simplifies to ${N \choose k} 2^{-N}$.

Since $\mathcal{H}_{1,2}$ and $\mathcal{H}(f)$ are nested,
then, assuming we have sufficiently many events,
we can use Wilk's theorem\footnote{%
  A potential issue in the case of $\mathcal{H}_1$ could be that the null value $f=0$ lies on the boundary
  of the domain~$[0,1]$ of~$f$, while Wilk's theorem requires the true value
  to be in the interior of the parameter space.
  However, $\ln(\mathcal{L}(k;f))$ has a well-behaved analytical continuation
  over a domain larger than $[0,1]$.
  As long as the estimator $\hat{f}$ has a sufficiently small variance,
  this boundary effect can therefore be ignored and Wilk's theorem still applies.
  See \cite{algeri_searching_2019} for a comprehensive discussion of the validity conditions of Wilk's theorem.
}
to try to exclude $\mathcal{H}_{1,2}$ .
To this end, we construct the two likelihood ratios $\Lambda_{1,2}(k)$ as:
\begin{equation}
  \Lambda_i(k) = \frac{\mathcal{L}_i(k)}{\mathcal{L}(k;\hat{f})}, \quad i = 1,2
\end{equation}
where $\hat{f}$ is the maximum likelihood estimator for~$f$:
\begin{equation}
  \hat{f} = \frac{1-a-k/N}{1-2a}
\end{equation}
Wilk's theorem states that if $\mathcal{H}_i$ ($i = 1 \text{ or } 2$) is realized,
then $-2\ln(\Lambda_i(k))$ follows a $\chi^2$ distribution with one degree of freedom.
Conversely, if we observe $-2\ln(\Lambda_i(k)) > 2.7$,
then $\mathcal{H}_i$ will be disfavoured at $90\%$~CL.
If both hypotheses $\mathcal{H}_{1,2}$ were disfavoured simultaneously,
this would suggest $\delta M \tau \sim 2\pi$ and potentially resolvable HNL oscillations.

If hypothesis~$\mathcal{H}_1$ is actually realized,
we expect $k$ to take a value around the expected number of events misclassified as LNV: $(1-a) N$,
which, for large~$N$, is approximately equal to the median.
The median of the log-likelihood-ratio when testing for $\mathcal{H}_2$ is therefore:
\begin{equation}
  \mathrm{med}_1\left(\ln(\Lambda_2)\right) \approx - N \underbrace{\big( \ln(2) + a\ln(a) + (1-a)\ln(1-a) \big)}_{\eqdef\; l_1(a) > 0}
\end{equation}
If, instead, $\mathcal{H}_2$ is realized,
then we expect $k$ to take a median value of approximately~$N/2$, such that:
\begin{equation}
  \mathrm{med}_2\left(\ln(\Lambda_1)\right) \approx N \underbrace{\left( \ln(2) + \frac{1}{2}\ln(a) + \frac{1}{2}\ln(1-a) \right)}_{\eqdef\; l_2(a) < 0}
\end{equation}
For a fixed confidence level,
we can invert these two formulas to estimate, for each true hypothesis~$\mathcal{H}_i$, $i=1,2$,
the median number of events~$N_i(a)$ required to exclude the other hypothesis:
\begin{equation}
  N_i(a) = \abs{\frac{\ln(\Lambda_{\mathrm{cr}})}{l_i(a)}}
\end{equation}
with $-2\ln(\Lambda_{\mathrm{cr}}) \approx 2.7$ for a $90\%~\mathrm{CL}$.
The higher the classification accuracy, the less events are required to reach the target,
while accuracies close to~$1/2$ do not allow distinguishing the two hypotheses, as $N_i(1/2) \rightarrow \infty$.
So far we have only considered the two extreme cases $f = 0 \text{ or } 1/2$, \ie{} $\delta M \tau \lessgtr 2\pi$.
We can generalize this analysis to the case where
the true hypothesis or the null hypothesis have a non-trivial LNV fraction~$f$.
A larger number of events will then be required to reach the same confidence level.
We will not discuss these cases further in this paper,
in order to avoid making the discussion unnecessarily complicated.

As a final step, for each HNL mass~$M$ and ratio $\abs{\Theta_e}^2 : \abs{\Theta_{\mu}}^2 : \abs{\Theta_{\tau}}^2$,
we compute the squared mixing angles $\abs{\Theta_{\alpha}}_i^2(M)$ required to produce $N_i(a(M))$ events,
thus producing for each true hypothesis~$\mathcal{H}_i$ a sensitivity limit,
above which \ship{} should be able to exclude the other hypothesis with a probability of at least~$1/2$.
The resulting sensitivity plots are presented in \cref{sec:results_lnv}.

\subsection{Systematic uncertainties coming from the heavy meson spectrum}
\label{sec:spectrum_systematics}

For a classifier to generalize well out of sample, \ie{} on real-world data,
the distribution used for training should match the true, physical distribution of features.
This is in general not the case, since a simulation never perfectly represents reality.
We can, however, work around this requirement by explicitly evaluating the classification accuracy
over a set of test distributions which is likely to encompass the true distribution.
This requires knowing and parametrizing the uncertainties coming from the simulation.
We can then obtain a conservative estimate for the classification accuracy by
varying the unknown parameters within their uncertainties, and taking a lower bound.
If this lower bound is high enough, we should still be able to probe lepton number violation on real data.

\begin{figure}
  \centering
  \includegraphics[width=0.65\linewidth]{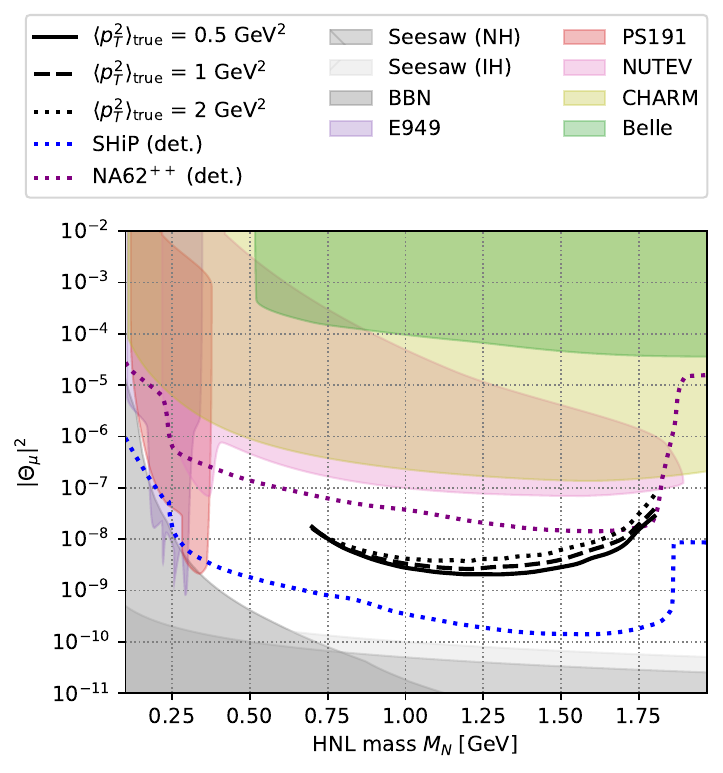}
  \caption{%
    Effect on the LNV sensitivity ($90\%$~CL) of computing the classification accuracy
    on a test set generated with a different $p_T$ spectrum compared to the training set,
    for an HNL coupling to the muon.
    Black lines represent the model-selection sensitivity of \ship{} for various true~$\braket{p_T^2}$.
    Here, the training set is always generated with $\braket{p_T^2} = \SI{1}{GeV}$.
  }
  \label{fig:spectrum_mismodeling}
\end{figure}

At \ship{}, the main uncertainty affecting the LNC / LNV classification accuracy
comes from the transverse momentum spread of the heavy meson spectrum,
which is only known with limited accuracy.
In order to estimate the actual sensitivity of \ship{} to LNV for a realistic dataset,
we therefore compute the classification accuracy for
a family of test sets generated using slightly different $p_T$~spectra,
and we take the lowest value as our estimate.
The change in the sensitivity resulting from varying $\braket{p_T^2}$
by a factor of two up and down with respect to the best-fit value from LEBC-EHS~\cite{lebc_d-meson_1987}
is shown in \cref{fig:spectrum_mismodeling}.
The planned charm spectrum measurements at \ship{} should be able to constrain
$\braket{p_T^2}$ to a much better accuracy than the range displayed in the figure.

Interestingly, when comparing this result with \cref{fig:pt_variation},
we observe that the classification accuracy seems to mostly depend on the $\braket{p_T^2}$ of the test set,
but not much on the one used for training.
This suggests that we might be able to safely use the best-fit spectrum for training
without worrying about biasing the results should the true spectrum turn out to be different,
provided that we use a conservative estimate for the accuracy.
In a more comprehensive study, one would likely want to vary additional parameters
related to the spectrum, geometry and simulation.

\bibliographystyle{JHEP}
\bibliography{bibliography}

\end{document}